\documentclass[prc,nofootinbib,noshowpacs]{revtex4-1}
\usepackage{graphicx,color}
\usepackage{amsmath}
\usepackage{adjustbox}
\usepackage[perpage]{footmisc}
\usepackage{multirow}
\usepackage{appendix}
\usepackage[nooneline]{caption}
\usepackage{url}
\usepackage{soul}
\newcommand{\beq}{\begin{equation}}
\newcommand{\eeq}{\end{equation}}
\newcommand{\beqn}{\begin{eqnarray}}
\newcommand{\eeqn}{\end{eqnarray}}
\usepackage[]{footmisc}

\begin{document}

\title{From bare to renormalized order parameter in gauge space:
structure and reactions}
\author{G. Potel}
\affiliation{National Superconducting Cyclotron Laboratory, Michigan State University, East Lansing, Michigan 48824, USA}
\author{A. Idini}
\affiliation{Department of Physics, University of Surrey, Guildford, GU2 7HX, UK}
\author{F. Barranco}
\affiliation{Departamento de F\`isica Aplicada III,
Escuela Superior de Ingenieros, Universidad de Sevilla, Camino de los Descubrimientos, 	Sevilla, Spain}
\author{E. Vigezzi}
\affiliation{INFN Sezione di Milano, Italy}
\author{R. A. Broglia}
\affiliation{Dipartimento di Fisica, Universit\`a degli Studi di Milano,
Via Celoria 16, 
I-20133 Milano, Italy}
\affiliation{The Niels Bohr Institute, University of Copenhagen, 
DK-2100 Copenhagen, Denmark}

\date{\today}

\begin{abstract}
The physical reason why one can calculate  with similar accuracy,
as compared to the experimental data, the absolute cross section
associated with two-nucleon transfer  processes between members of pairing rotational
bands, making use of simple BCS (constant matrix elements) or of many-body 
(Nambu-Gorkov (NG), nuclear field theory (NFT)) spectroscopic amplitudes, is not immediately obvious. 
Restoration  of spontaneous symmetry breaking and associated emergent generalised rigidity in gauge space provides  the
answer, and points to a new emergence: a physical sum rule resulting 
from the intertwining of  structure and reaction processes  
and closely connected with the central role induced pairing interaction
plays in structure together with 
the fact that  successive
transfer dominates  Cooper pair tunnelling.  
\end{abstract}

\pacs{
 21.60.Jz, 
 23.40.-s, 
 26.30.-k  
 } 
 \maketitle
\date{today}

\section{Introduction}

The starting point of most descriptions of nuclear structure and reactions is based on independent particle motion.
The validity of such a picture is related to basic quantum mechanics. Potential energy privileges
fixed position between particles. Fluctuations, in particular quantum fluctuations, the only ones operative in a nucleus
in its ground state, symmetries. Regarding single-particle motion, such competition is embodied in the quantality parameter 
\cite{Mottelson1998} ,
\begin{equation}
q = \frac{\hbar^2}{m a^2} \frac{1}{|v_0|},
\end{equation}
where $m$ is the nucleon mass, $v_0$ and $a$  being the strength and the range of the strong NN-potential 
respectively ($v_0 \approx - 100 $ MeV, $a \approx$  1 fm). The
above equation is the ratio between the kinetic energy of confinement and the potential energy. Because $q \approx 0.4$, nucleons in the nucleus are delocalized,
and mean field is a good approximation. In particular, the HF mean field.

\section{Spontaneous symmetry breaking}

The fact that basic properties of a quantal system can be described in terms of a mean field solution  which does not display some of the symmetries of 
the original Hamiltonian is the spontaneous symmetry breaking phenomenon. 

The lower symmetry mean field solution  defines a privileged orientation in the corresponding three-dimensional
(e.g. Nilsson) , gauge (e.g. BCS, HFB) , etc. space.  All orientations have the 
same energy, in keeping with the fact that  the restoring constant associated with changes  in the 
Euler-, gauge-, etc. angles is zero. Fluctuations 
in orientation  thus diverge  in precisely the right  manner
to restore symmetry (see e.g.  \cite{Brink2005}, Sects. 4.2. and 4.2.3 and refs. therein). Because this divergence is associated with the vanishing of 
the frequency for constant inertia, the system acquires  generalised rigidity
(emergent property). Thus, acting with the specific external field (Cooper pair transfer in the case of pairing rotational band), sets the deformed system into rotation as a whole, without retardation effects. 
The above phenomena are at the basis of the broken symmetry restoration paradigm used to identify  the elementary modes of nuclear excitation
(see e.g.  \cite{Bohr1975} and refs. therein).
In particular pairing rotations \cite{Bohr1975,Bes1966,Broglia2016,Broglia2000,Hinohara2016,Lopez:13}.

Pairing in nuclei  has been introduced a number of times. The first to explain the enhanced stability of even as compared to odd nuclei \cite{Heisenberg1932}. 
Subsequently, to describe 
the correlations associated  with such staggering  effects \cite{Mayer1950,Racah1952}. After the BCS explanation of  superconductivity \cite{Bardeenetal1957a,Bardeenetal1957b}, to account 
for the presence of a gap in the low-energy intrinsic  excitation spectrum of deformed nuclei  \cite{Bohr1958}.
Finally, in connection with the advent of  the Josephson effect, namely Cooper pair tunnelling, and the study of two--nucleon transfer processes, specific probes of 
deformation in gauge space  \cite{Bohr1964,Yoshida1962}.

\subsection{Order parameter of nuclear superfluid phase }

The order parameter  associated with  independent pair motion is defined as, 

\begin{eqnarray}
\alpha'_0 = \langle BCS(N+2)| P '^+  |BCS(N)\rangle  ,   \nonumber \\
=\sum_j \sqrt {\frac{2j+1}{2}}  B(j^2(0), N \to  N+2).
\label{eq:alpha0x}
\end{eqnarray}
That is, the number of pairs participating in the BCS condensate. The quantity
\begin{eqnarray}
B(j^2(0), N \to  N+2) = \langle BCS(N+2)| T '^+(j^2(0))|BCS(N)\rangle  , \nonumber  \\
= \sqrt{ \frac{2j+1}{2}} U'_j(N) V'_j(N+2),
\label{eq:Bj}
\end{eqnarray}
is the two-nucleon transfer spectroscopic amplitude,
\begin{equation}
T '^+(j^2(0)) = \frac {[{a'_j}^+ {a'_j}^+]_0} {\sqrt{2}},
\label{eq:Tj}
 \end{equation}
 being the two-nucleon (Cooper pair) transfer operator, while
\begin{equation}
P '^+  =\sum_{jm>0} {a'_{jm}}^+ {a'_{\bar{jm}}}^+ = \sum_j \sqrt{ \frac {2j+1}{2}} T ' (j^2(0)),
\end{equation}
is the operator which creates a pair of particles in time reversal states. $|BCS(N)\rangle  $ 
labels the BCS state for which the $\lambda$ parameter 
(Fermi energy) has been adjusted so that 2$\sum_{jm>0} {V'_j}^2 = N. $

In keeping with (\ref{eq:alpha0x}), the order parameter $\alpha'_0 = \sum_j  \left( \frac{2j+1}{2} \right) U'_j(N) V'_j(N+2) \approx \sum_j   \left( \frac{2j+1}{2}   \right) U'_j V'_j$ provides a
 measure of the nuclear deformation in gauge space, and thus of the fact that the system displays a privileged orientation 
in this space, as can be seen from the relation (see App. A) 
\begin{equation}
\alpha'_0= \sum_j  \left(\frac{2j+1}{2}\right)  U'_j V'_j = e^{2 i \phi} \sum_j \left( \frac{2j+1}{2} \right)  U_j V_j =
e^{2i \phi} \alpha_0,
\label{alpha0}
\end{equation}
where  the primed quantities are the BCS occupation amplitudes referred to the intrinsic system of reference in gauge space (i.e. body-fixed
BCS state), while the unprimed quantities are the same quantities referred to the laboratory system of reference. 
The two systems are connected by a rotation in  gauge space of angle $\phi$, induced by the operator 
${\cal G} = exp(-i \hat N \phi)$, $\hat N$ being the 
number operator and thus $a'^+_{jm} = {\cal G}(\phi) a^+_{jm} {\cal G}^{-1} (\phi)$ (see e.g. \cite{Poteletal2013} and refs. therein)\footnote{In the remaining of this paper, although we continue to refer all quantities to the intrinsic,body-fixed frame of reference in gauge space, we will not use primed letters, exception made in particular cases which will be signaled, and where the  explicit appearance of the gauge angle 
$\phi$ is of use (cf. e.g. App. A, Eq. (\ref{BCSphi})).}.  

A simple empirical confirmation that $\alpha_0$ is  the number of  Cooper pairs of a superfluid nucleus can be made with the help of the single $j-$shell model.
In this model  $V_j = (N/2 \Omega)^{1/2}$ and $U_j = (1 - N/2\Omega)^{1/2}$, where
$\Omega= (2j+1)/2$. For a system with $N=\Omega$ particles, i.e. $\Omega/2$  pairs, half filled shell, typical of  a superfluid nucleus, $V_j = U_j =(1/2)^{1/2}$ and $\alpha_0 = \Omega/2$. Thus, $\alpha_0$ gives an estimate of the number of Cooper pairs which participate  in 
specifying  the orientation the $|BCS\rangle  $ state has in gauge space.  With the help of  the approximate expression
 $\Omega = (2/3) A^{2/3}$ one obtains, for $^{120}$Sn, $\alpha_0 = 8$. Detailed microscopic calculations
 give values of $\alpha_0 = 5-6$ (see Sect.\ref{Section6} Table \ref{table3x}). 
 
Symmetry restoration results from zero point fluctuations of the gauge angle  setting 
the BCS deformed state into rotation and leading to pairing rotational bands, e.g. the ground state 
of superfluid Sn-isotopes, where $N$ plays, in gauge space, the role angular momentum plays in 
quadrupole rotational motion. This symmetry restoration can be implemented by diagonalizing in QRPA the residual
interaction $H_{res}$ acting among quasiparticles and neglected in the BCS mean field approximation (cf. App. A).

Because there are two parameters which determine  the admixture of particle and hole states connected with gauge symmetry breaking, namely $U_j$ and $V_j$ (quasiparticle transformation), there are only two fields $F$ which contribute to $H_{res}$  through terms  of the type $FF^+$. One, antisymmetric with respect to the Fermi energy, namely $U_j^2 - V_j^2$ and leading to pairing vibrations of the gauge deformed state $|BCS\rangle  $ ($H'_p$ contribution to $H_{res}$, cf. e.g.  \cite{Brink2005}, App. J). The other one,  ${U_j}^2 + {V_j}^2$ is symmetric with respect to $\epsilon_F$ and leads to fluctuations which restore gauge symmetry ($H''_p$ contribution to $H_{res}$,  $H_{BCS} + H''_p$ commute with $\hat N$). Within this scenario, the field $U_j^2 - V_j^2$ excites two-quasiparticle states. Eliminating  (in a particle-conserving fashion) this contribution from $U_j^2+ V_j^2$, one obtains the field which connects the members of a ground state rotational bands. 
That is $((U^2_j+ V^2_j)^2 - (U^2_j - V^2_j)^2)^{1/2} \sim U_j V_j$.
This result, together with (\ref{eq:Bj})  and (\ref{eq:Tj}), testifies  to the fact that  two-nucleon transfer reactions are, from the point  of view of structure,  the specific probes of pairing condensation in nuclei \cite{Poteletal2013a}, as it emerges in a natural fashion writing
\begin{equation}\label{eq7}
\alpha_0=\langle BCS|\sum_j\left(\frac{2j+1}{2}T(j^2(0))\right)|BCS\rangle
\end{equation}

It is then natural that\footnote{Within this context one is reminded of the fact that the Coulomb excitation cross section associated with the excitation of members of a quadrupole rotational band is proportional to $Q_0^2$, the square of the quadrupole moment providing a measure of the number of aligned nucleons \cite{Bohr1975}} the absolute two-nucleon transfer cross section  between members of a pairing rotational band can, schematically,  be written as 
\begin{equation}
\sigma \sim |\alpha_0|^2,
\label{eq:sigma}
\end{equation}
 emphasizing again the close connection (unification) of structure and reaction aspects of the subject under discussion.

\section{Physical nucleons and induced pairing}\label{s3}

\begin{figure}[h!]
\centerline{\includegraphics[width=0.8\textwidth,angle=0]{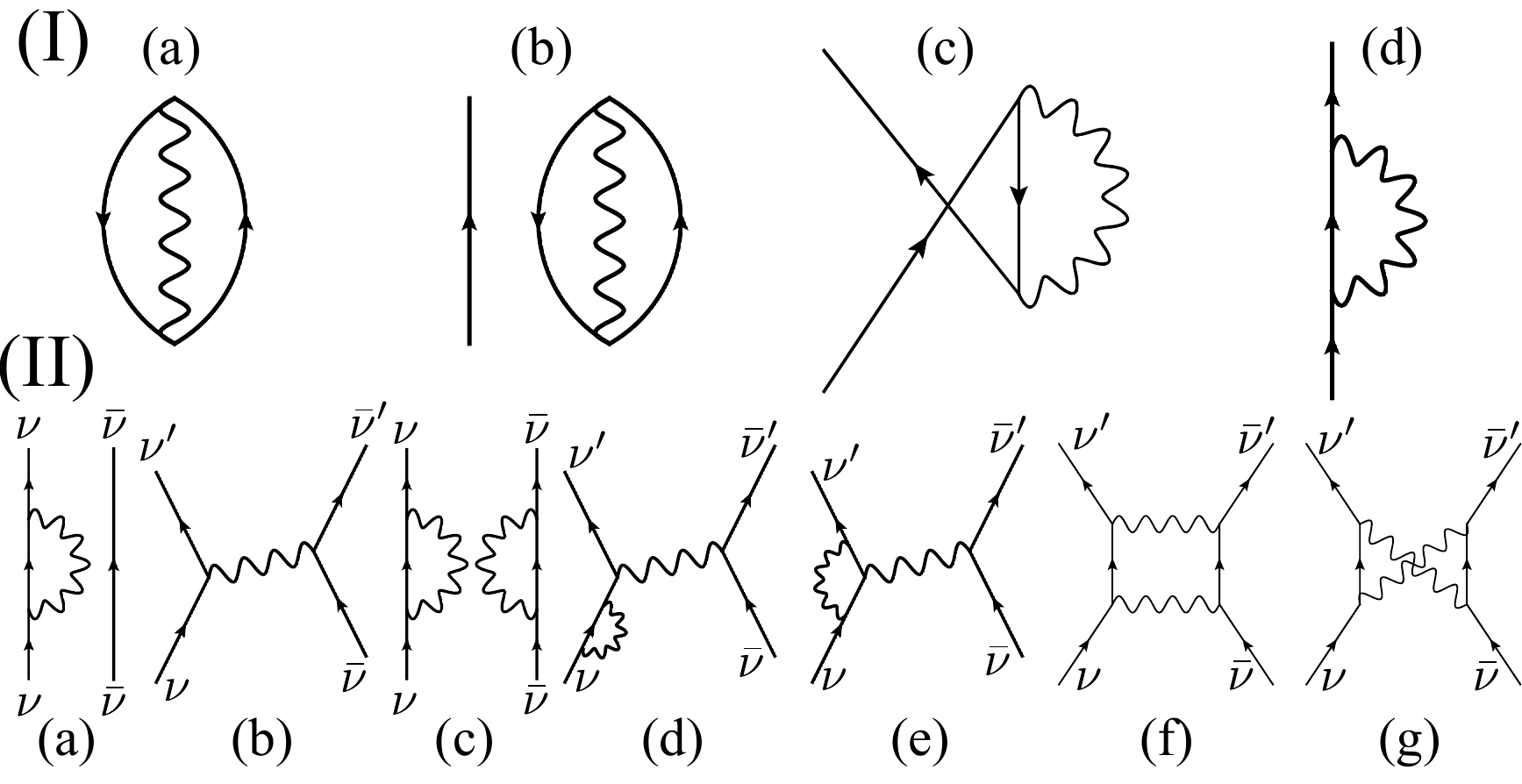}}
\caption{
{\bf (I) (a)} ZPF associated with (particle-hole) surface vibrations; 
{\bf (b)} odd system; {\bf (c)} the antisymmetrization between the particles  considered explicitly and 
those involved in the vibration; {\bf (d)} time ordering of (c). 
Diagrams (c) and(d) lead to the clothing of single-particle motion  in lowest order in the 
particle-vibration coupling vertex.
{\bf (II)} A dressed  nucleon moving in a state $\nu$ in the presence of: {\bf (a)} a bare nucleon
moving in the time reversed state $\bar \nu$,
{\bf (c)} another dressed nucleon. Exchange of vibration in $(a)$ leads to {\bf (b)}, the NFT 
lowest-order contribution in the particle-vibration coupling vertex, of the induced pairing interaction (App. G). Exchange 
of vibrations in {(c)}  leads to  {\bf (d)} self-energy, {\bf (e)} vertex correction of the  
induced pairing interaction (App. H);  {\bf (f)} ladder diagram contributing to the induced pairing interaction.
The symmetrisation between the bosons displayed in (c) is shown in {\bf (g)}.}\label{fig2}
\end{figure}

In what follows we will show  that there is a simple physical reason at the basis of the above parlance, rooted  on the fact that the atomic nucleus is a 
leptodermous finite many-body quantal system. Virtual states, like those associated with zero-point fluctuations (ZPF) of the nuclear vacuum (ground state), e.g, in which a surface quantised vibration and an uncorrelated particle-hole mode get virtually excited for a short period of time (Fig. \ref{fig2}, I(a)) are a basic characterising feature of these systems \cite{Baroni} .  Adding 
a nucleon to it (odd system, Fig. \ref{fig2} (I) (b)) leads, through the particle--vibration coupling strength ($V$, (see e.g. \cite{Schuck} Eq. (C6)))  to processes which contain the effect of the antisymmetry between the single-particle explicitly considered and the particles out of which  the vibrations are built  (Fig. \ref{fig2} (I)(c)). Time ordering 
gives rise to the graph shown in Fig. \ref{fig2} (I)(d). Processes I(c) and I(d)  known as correlation (CO) and polarisation (PO) contributions to the mass
operator (see \cite{Mahaux} and refs. therein)  cloth the particles, leading to physical nucleons  whose properties can be compared with the experimental findings. Summing up, 
the processes shown in Fig. \ref{fig2} (I) are textbook examples of quantal nuclear phenomena testifying to the fact that the clothing of nucleons is at the basis of the quantal description of the atomic nucleus. 

Nuclear superfluidity at large, and its incipience in the case of single Cooper pair like e.g. in $^{11}$Li in particular, are among the most quantal of all  the phenomena displayed by the nuclear many-body 
system. Even if the $^1S_0$, $NN-$interaction was not operative, or was rendered subcritical by screening effects as in the case of $^{11}$Li, Cooper binding will still be healthy, as a result of the exchange of vibrations between pairs of physical (clothed) nucleons moving in time reversal states close to the Fermi energy (Fig. \ref{fig2} (II)(b),(d)-(g)), a direct
consequence of the ZPF of the  nuclear vacuum (ground state) (Fig. \ref{fig2} (I)(a) and (c))
\footnote{Within this context,  let us note that the Hamiltonian contains in the potential
energy, the classical idea of force, the Newtonian quantitative expression for causation. If, for instance,
particles are acting on one another with a Coulomb force 
(as protons in the nucleus or the nucleus and the electrons in an atom), there appears in $H$ the
same timeless action over finite distance as  in Newtonian mechanics. These vestiges of classical causality
can give rise to serious problems under certain circumstances (cf. e.g. \cite{Born,Schwinger,Feynman} ), problems which are 
eliminated by taking into account the fact that the Coulomb interaction arises from the exchange of 
 photons  between charged particles.
It is interesting to quote from the notes of Feynman on the self-interaction of two particles: "... the self energy 
of two electrons is not the same as the self-energy of each one separately. That is because among the intermediate 
states which one needs in calculating the self-energy of particle number 1, say, the state of particle 2 can no
longer appear in the sum because a transition of 1 into the state of 2 is excluded by the Pauli 
exclusion principle. The amount by which the self-energy of two particles differs from the self-energy of each one separately is actually  the energy 
of their electric attraction."

Fluctuations, in quantum mechanics, not only enter through the kinetic energy, but also through the
potential energy.
Because of Heisenberg's indeterminacy relations and Born-Jordan commutation laws,  the quantal many-body
system even in its ground  state is at a finite, effective "temperature", and the separation 
between enthalpies  (potential) and entropic (kinetic) components is not clear cut, as forces are also $\omega-$dependent  
many-body phenomena which only approximately 
can be treated in terms of static terms.

In other words, the central issue in the quest of solving 
the many-body problem is that of having a correct description  of the ground state
as far as it reflects the virtual excitation of the system (App. D). This is the reason why
effective field theories in general and NFT in particular have  a good starting point, while 
{\it ab initio }  calculations have to create it at each stage. While this task may not be too complicated 
to describe the effect of ZPF associated with giant resonances, that associated with  low-lying 
collective modes is likely more trying. On the other hand, these states play  the dominant role,
through their state dependent ZPF, in determining the  texture of the nuclear physical vacuum. }

Within this context, and only so, is that one can posit that the order parameter $\alpha_0$ does not depend on the presence or less
of  the $^1S_0$, $NN-$bare potential. 
Independent Cooper pair motion and thus nuclear superfluidity is intrinsically contained in the fluctuations  of the quantal nuclear vacuum. As such, 
it is a truly emergent many-body nuclear property  implying generalised rigidity in gauge space, the associated pairing rotational bands 
being specifically excited through pair transfer \cite{Bes1966,Broglia2000,Hinohara2016}. 
The fingerprint of spontaneous symmetry breaking in finite many-body systems is the presence of rotational bands  associated with symmetry restoration. 
To qualify as a rotational band, a set of levels must display enhanced transition probabilities (absolute
cross sections), associated with the operator  having a non-vanishing value in the (degenerate) ground state (order parameter). In the present case
(pairing rotational bands), of the two-nucleon transfer operator \cite{Poteletal2013,BrogliaHansenRiedel}. In other words, cross talk (absolute transfer cross sections) between a member of a pairing
rotational  band and states not belonging to it, should be much  smaller than between members of the band. It could be argued that also important for the characterisation 
of a pairing rotational band is the parabolic dependence of the energy with particle number. True, but many non-specific aspects can modify this dependence, without altering the gauge  kinship (common $|BCS\rangle   $ like  intrinsic state).
 Before adscribing to two-nucleon transfer processes the role of specific probe not only 
from the point of view of nuclear structure, but also from the vantage point  of nuclear reactions \footnote{Within this context one may mention that while
e.g. $(p,t)$ reactions are quite attractive processes to learn about pairing in nuclei, the $s-$relative motion of
the two transfer nucleons is quite different in the target nucleus, e.g. $^{120}$Sn, than in the outgoing  triton ($\Omega_n$ overlaps, \cite{BrogliaHansenRiedel}).
Inducing Cooper pair transfer with heavy ions allows to better probe the $s$--correlations. On the other hand, the simultaneous 
opening of many other channels makes the analysis of such reactions more involved and, arguably, less reliable.}, 
a subject concerning the 
reaction mechanism is to be clarified. 
\newpage

Making use of the fact that in superfluid nuclei lying along the stability valley like e.g. the Sn-isotopes, about half of the neutron pairing gap is associated with the induced pairing interaction \cite{EPJ,Idini2015}, that is, $\Delta_{ind} \approx g_{pv} \alpha_0 = \Delta_{exp}/2 \approx 0.8 $ MeV, where $g_{pv}$ is the particle-vibration coupling parameter (equal to minus the induced pairing interaction), and of the fact  that the mass enhancement factor $\lambda$ (i.e. $m_{\omega} = m(1+\lambda) \approx 1.4 m$, see e.g. \cite{Mahaux}, see also \cite{Schuck} Eq (C11)) can be written as $\lambda= g_{pv} N(0) (\approx 0.4)$, one obtains (see App. E)
\begin{equation}
N(0) \approx  \frac{1}{2} \alpha_0 \; {\rm MeV}^{-1} \approx 4 \; {\rm MeV}^{-1},
\label{eq:N0}
\end{equation}
for the density of neutron levels of e.g. $^{120}_{50}$Sn$_{t0}$ at the Fermi energy and for one spin orientation, as experimentally observed
($a \approx   N/8$ MeV$^{-1}$ for both spin orientation, see  \cite{Bortgdr}, Eq (7.16)).

As a consequence of Eqs. (\ref{eq:sigma}) and (\ref{eq:N0}),
\begin{equation}
\sigma \sim (N(0))^2.
\end{equation}
In other words, Cooper pair tunneling in nuclei is dominated by successive transfer. The significance of this result  becomes clearer by recalling the fact that according to the golden rule, to processes like tunneling or decay is associated a decay width linear in the density of states. 

One can then argue that successive transfer may imply pair breaking,  making two-nucleon transfer reactions a less than ideal
probe of pairing correlations in nuclei. That this is not so can be understood by calculating  the correlation length, 
that is, the range over which Cooper pairs partners,
 correlated by the exchange of collective vibrations, feel the presence of each other. 
 One obtains 
\footnote{In keeping with the fact that $\Delta_{ind} \approx \Delta_{exp}/2$ as stated above, the actual Cooper pair mean square 
radius of $^{120}$Sn, i.e. the order parameter, is about half the value (\ref{xivalue}).}
(see App. F)
\begin{equation}
\xi_{ind} = \frac{\hbar v_F}{\pi \Delta_{ind}} = \frac{\hbar v_F}{\pi g_{pv} \alpha_0 } = \frac{\hbar v_F N(0)}{\pi \lambda \alpha_0} \approx 24 \; {\rm fm}.
\label{xivalue}
\end{equation}
On carrying out the above estimate use has been made of $v_F /c = 0.3$, $\lambda = 0.4$ and $\alpha_0 =8 $. 
As a result, the generalised pair quantality  parameter

\begin{equation}
q_{\xi_{ind}} = \frac{\hbar^2}{(2m) \xi_{ind}^2} \frac{1} {g_{pv} \alpha_0} \approx
\frac{\hbar^2}{2m \xi_{ind}^2} \frac {N(0)}{\lambda \alpha_0} \approx 0.04,
\label{quant1}
\end{equation}
has a value much smaller than 1, implying potential energy dominance and thus  to a strong correlation of the two
partner nucleons of the Cooper pair over distances   of the order of $\xi$, quantity larger than nuclear 
dimensions.
This result
testifies to the fact that  successive transfer of nucleons fully probes the nuclear pair correlations.

The wave function of the nucleons in the pair are phase--coherent, so one has to add
the transfer amplitudes before taken the modulus  square. The nucleons do not tunnel independently, but more like a single particle, and
the probability  of a pair going through is comparable to the probability  for a single nucleon. It is like interference 
in optics with phase-coherent wave mixing. In a nutshell, and denoting 
$P_1$ and $P_2$ the single and pair nucleon transfer probability \footnote{Single-particle $^{120}$Sn(p,d)$^{119}$Sn and two-nucleon
transfer $^{120}$Sn(p,t)$^{118}$Sn(gs) absolute cross sections are in both cases, of the order of  few mb (see \cite{Idini2015} and refs. therein).}, one can write
\begin{align}\label{eq13}
\nonumber P_2 &=  \left| \frac{1}{\sqrt{2}} \left( e^{i \phi'} U\sqrt{P_1} + e^{i \phi}V \sqrt{P_1} \right)\right|^2 = P_1 \frac{ (1+ 2UV\cos \epsilon)}{2}\approx P_1\\
&\quad (\epsilon \equiv \phi - \phi'),
\end{align}
 where the assumption was made that $\epsilon=0$ and $U=V=1/\sqrt{2}$, in keeping with the fact that
the  ``single particle'' pair wavefunction is $(U_{\nu} + V_{\nu} e^{- i 2 \phi} a^+_{\nu} a^+_{\bar \nu})|0\rangle  $, and of the single--$j$ shell estimate of the BCS occupation amplitudes.

\begin{figure}[h!]
	\includegraphics[width=0.7\textwidth]{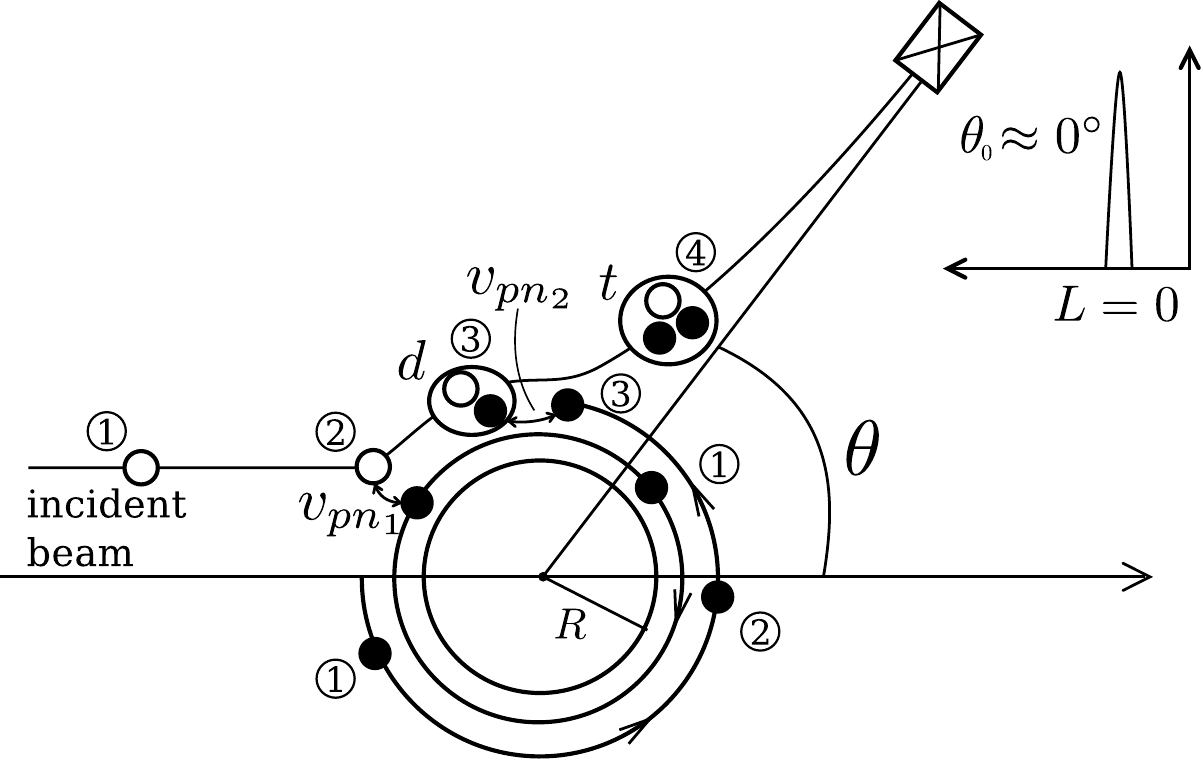}
	\caption{Schematic representation of Cooper pair transfer in the reaction $^{120}$Sn$(p,t)^{118}$Sn (gs) leading to essentially a single peak in the spectrum at $0^\circ$ (after \cite{Poteletal2013a} Fig 1 (II) (b) and (c)).}\label{figscripta}
\end{figure}

Summing up, in the reaction $^{120}$Sn +p $\to$ ($^{119}$Sn + d) $\to$ $^{118}$Sn + t, the first neutron 
of the Cooper pair picked up  by the proton to constitute the (virtual) deuteron can be at the surface of the nucleus close to
the proton, while the second one can be at the antipode (diameter $\approx$ 12 fm), eventually the second one being transferred 
to form the triton within the interaction range ($\approx $ 2 fm). This scenario involves relative distances between the partners 
of a Cooper pair, one in the target the other one in the (virtual) deuteron, of the order of 10-14 fm (see Fig. \ref{figscripta}, where the inner $(n_1)$
orbital motion is to be interpreted to schematically describe clockwise motion, the external one ($n_2$), anticlockwise one). Thus transfer of a rather extended object made out of two neutrons moving in time reversal states, still correlated as a single--particle 
of mass $2 m^*$, in keeping with the estimated value of $\xi$ and of the phase coherence expressed by the relation $P_2 \approx P_1$.

Following practice we refer throughout this paper to structure and reactions as two separate issues in the study of the atomic nucleus. While likely pedagogic, such an approach is fundamentally wrong as already suggested in the abstract. In a nutshell, structure and reactions are two aspects of the same subject. One likely involving bound the other continuum states, a distinction which is not even operative universally, certainly not in the case of light exotic halo nuclei. But more important, because in quantum mechanics one can hardly call physical a non measurable feature of a system.

Within this context, the quantity (\ref{alpha0}) modulus squared cannot be measured, but only when each of its terms are properly weighted by the formfactors (i.e. successive as well as simultaneous  and non--orthogonality functions), and energy denominators (Green functions), as forcefully expressed in (\ref{eq7}) (see also \cite{Poteletal2013a}, App A in particular Eq. (A.21), \cite{Poteletal2013} Sect. III specially Eq. (38 (b)) as well as \cite{Physica_scripta}, Figs. 2 and 12). Consequently, when discussing about the order parameter $\alpha_0$, in particular concerning the possible emergence of a physical sum rule, we are all the time aware of this fact even if, for simplicity, we do not state it explicitly. In other words, talking about $|\alpha_0|^2$ the ultimate reference is to the results displayed in Figs \ref{fig:cross} and \ref{fig5}, namely predicted observables (absolute differential cross section) in comparison with the experimental findings. This implies that each term of $\alpha_0$ has to be viewed as the weighted $j^2(0)$, mainly successive, formfactor associated with independent pair motion, in a similar way in which, talking about one--nucleon transfer, independent particle motion implies a spectroscopic amplitude and a radial formfactor, also renormalized if that is the case (see e.g. \cite{11Be} and refs. therein). While the results contained in Table \ref{table_novert2} and \ref{table3x} play an important role in the calculation of observables, the different entries still refer to the assessment of theory against theory.

With the above proviso we can state that in keeping with  the fact that  $|BCS\rangle  $ is a coherent state,  displaying  off diagonal long range order\footnote{Within this context, it is of notice that the overall gauge phase ensuring that $|BCS\rangle$ is a coherent state in this space, is the same as the one at the basis of the Josephson effect. In fact, the Josephson effect provided the first (only) specific probe to measure the gauge angle (difference) in superconductors. Now, because in condensed matter there are a number of phenomena like supercurrents, Meissner effect, etc., which testify to pair condensation, the direct relation existing between ODLRO and Josephson effect has not been at center stage. However, the situation is completely different in the case of atomic nuclei, where supercurrents cannot be observed, in keeping with the fact that $\xi\gg R_0$. Consequently, Cooper pair transfer is essential to probe nuclear superfluidity.} (ODLRO, see Apps. A, B and C), 
one expects (\ref{eq:sigma}) to be a physically conserved quantity. Also that the robustness of the order parameter $\alpha_0$ to characterise nuclear superfluidity as compared to the pairing gap is testified by the fact that $\alpha_0$ is different from zero also in nuclear 
regions, like between two heavy ions at the distance of closest approach in e.g. the process
$a (= b+2) + A \to b + B (=A+2)$, situation in which the pairing interaction and thus also $\Delta$ are zero\footnote{Using an analogy, the deformation of a 3D--quadrupole--rotating system is measured by the quadrupole moment $Q_0$, and not by the field approximation ($\kappa Q_0$) to the separable quadrupole--quadrupole interaction $H_Q=-\kappa(Q\cdot Q)$.}. 

Let us conclude this Section by noting that while the expression (\ref{eq13}) displays in a simple way the gauge phase coherence associated 	with independent pair motion, it does not contain the independent particle limit, lacking the energy denominator. This limit is of course simple to exhibit in the quantal \cite{Poteletal2013} or semiclassical  \cite{Poteletal2013a} formalism mentioned above, which has the backdraw of becoming involved in connection with phase gauge coherence.
   
\section{Many-body aspects of the nuclear pairing interaction}

While in condensed matter the many-body aspects of the pairing interaction could not be ignored,  this could happen in nuclear physics.
This is primarily due to the fact that  the electron-electron
bare interaction is repulsive (Coulomb). But also because of the fact that the highest values of $T_c$ in low-temperature metallic superconductors 
are, as a rule,  associated with bad conductors at  room temperature , underscoring the role played by the electron-phonon coupling
in the superconducting phenomenon, and the need for a correct treatment of this interaction. In other words, the scenario of the Nambu--Gorkov and Eliashberg approach to superconductivity \cite{Gorkov,Nambu,Eliashberg}.

In nuclear physics, on the other hand,
 the values of $^1S_0$ phase shifts
are positive  for low values of the relative nucleon  velocities ($E_{lab} \leq $ 200 MeV), let alone the fact that 
the particle-vibration coupling mechanism is still often thought to give only rise to self-energy phenomena.
As a result, it was assumed that the nuclear pairing interaction was short range 
and resulting solely from meson exchange, long range interactions being responsible for mean field effects
(see e.g. \cite{Bessorensen} and refs. therein), an attitude which has proven to be difficult to overcome. In other words, similarly to the fact that one cannot measure the bare nucleon mass in  nuclei but the clothed one (see Fig. \ref{fig2} (I)(c),(d)), 
one cannot measure the  bare pairing interaction  in the nuclear medium but the effective one, 
sum of the bare ($v^p_{bare}$) and of the induced ($v_p^{ind}$) one (see Fig. \ref{fig2} (II)(b),(d)-(g)). Furthermore, in nuclear physics as in condensed matter, a non--perturbative treatment of the PVC is needed in a number of cases, e.g. in connection with the breaking of the $d_{5/2}$ orbit of $^{120}$Sn.

Applying, within the framework of NFT,  
the Nambu-Gorkov technique developed to describe metallic superconductors to this open shell nucleus, it is possible
 to obtain a complete characterization of it. The theoretical predictions reproduce the experimental results  within the 10\% level \cite{Idini2015}.
As we shall see below, the contributions of the  many-body effects related to the one-particle channel  do not affect 
 the absolute two-nucleon transfer reaction cross section in any major way. This fact
 testifies  to the robustness of $\alpha_0$, in the sense of two--nucleon transfer spectroscopic amplitude as explained in Sect. \ref{s3},  and to the physical soundness to make it 
the nuclear superfluid order parameter.

\section{Elementary modes of excitation: empirical renormalization  in structure and reactions}

The elementary modes of excitation  of a many-body system  represent a generalization of  the idea of normal modes of vibration.
They constitute the building blocks of the excitation spectra, providing  insight  into the  deep nature  of the system one is studying, aside from allowing 
for an economic description  of complicated spectra in terms of a gas of, as a rule, weakly interacting bosons and fermions. In the nuclear case 
they correspond to clothed particles and empirically renormalised vibrations (rotations).

There lie two ideas behind the concept of elementary modes of excitation. First, that one does not need to be able to calculate the total binding
energy  of a nucleus to accurately describe the low energy excitation spectrum, in much the same way in which one can calculate 
the normal modes of a metal rod not knowing how to  calculate its total cohesive energy.
The second idea is that low-lying states ($\hbar \omega \ll \epsilon_F \ll BE)$ are of a particularly simple
character, and are amenable to a simple treatment, their
interweaving  being carried out at profit, in most cases,  in perturbation theory
\footnote{More precisely, and in keeping with  the fact that 
boson degrees of freedom have to decay through linear particle-vibration 
coupling vertices into their fermionic components to interact with another vibrational mode,
the interweaving between the variety of many-body components clothing a single-particle state 
or a collective vibration will be described at profit in terms of an arrowed matrix which, assuming perturbation theory
to be valid, can be transformed, neglecting contributions of the order of $g^3_{pv}$ or higher, into a co-diagonal matrix, namely a matrix 
whose non-zero elements are $(i,i-1)$ and $(i,i+1)$,  aside from  the diagonal ones $(i,i)$.}. 
Within this context it is  necessary to have a microscopic description 
of the ground  state of the system  which ensures that it acts as the vacuum state 
$|\tilde0\rangle  $ of the elementary modes of excitation. In other words $a_{\nu}|\tilde 0 \rangle   = 0, \Gamma_{\alpha} |\tilde 0\rangle   =0$, where
$a^+_{\nu}|\tilde 0 \rangle   = |\nu\rangle  $ and , $\Gamma^+_{\alpha} |\tilde 0\rangle   =|\alpha\rangle  $ represent a single-particle and a one-phonon state.
This   implies, in keeping 
with the indeterminacy  relations $\Delta x \Delta p \geq \hbar/2$, that $|\tilde 0\rangle   = |0\rangle  _F |0\rangle  _B$
displays quantal zero point fluctuations (ZPF).

Within the framework of nuclear field theory (NFT) used below, in which single-particle (fermionic, F) and vibrational
(bosonic, B) elementary modes of excitation are to be calculated within the framework of HFB and QRPA
respectively, $|\tilde 0\rangle  $ must display the associated ZPF (cf. App. D). In particular for (harmonic) vibrational modes
$\Delta x \Delta p = \hbar/2$, the associated zero point energy amounting to $\hbar \omega/2$
for each degree of freedom, e.g 5$\hbar \omega/2$ for quadrupole vibrations, 
$\hbar \omega$ being the energy of the collective vibrational mode under consideration.

An illustrative example of the above arguments is provided by the low-lying quadrupole vibrational state of $^{120}$Sn. 
Diagonalizing SLy4 in QRPA leads to a value of $B(E2)$
(890 $e^2$ fm$^2$) which is about a factor of 2 smaller than experimentally observed (2030 $e^2$ fm$^2$). 
Taking into account  renormalisation effects in NFT, 
namely in a conserving  approximation (self-energy and vertex corrections, generalised Ward identities), one obtains a value (2150 e$^2$ fm $^2$), 
which essentially coincides with 
the experimental findings. One does not know how to accurately calculate the absolute ground state energy 
$E_0$ (total binding energy) of e.g. $^{120}$Sn, but one can do pretty well to work out 
the properties of the low-energy mode of this nucleus, also the collective energies 
$\hbar \omega_L =E_L - E_0$, and thus the associated ZPF and zero point energy $E_0$,
by renormalizing  QRPA solutions to  lowest order through self-energy and vertex corrections contributions  \cite{EPJ}. 
Now, if the collective phonons are not the main object  of the study, but are to be used to cloth the single-particle states 
and give rise to the induced pairing interaction, one can make 
use of phonons which account for the experimental findings (empirical renormalization \cite{Idini2015}, see also \cite{Physica_scripta,11Be}).

It is to be noted that in calculating the $E \lambda$ lifetimes , e.g. the quadrupole lifetime associated with the low-lying quadrupole mode 
$(T(E2) = 1.22 \times 10^9  \times E_{\gamma}^5 \times
B(E2), \;  E_{\gamma} = \hbar \omega_{2^+}$), the kinematic ($E_{\gamma}^5)$ and structure ($B(E2)$) contributions can be treated separately.
This is in keeping with the fact 
that in the case of electromagnetic decay as well as of  anelastic processes, the relative motion coordinate is always that of the entrance channel, at variance with particle transfer processes. Consequently, in connections with these processes,
 structure and reactions are  treated separately, a possibility not operative in the case of transfer reactions.
 Let us extend this discussion to  particle transfer process. In particular, to the two-particle pickup reaction $^{120}$Sn(p,t)$^{118}$Sn(gs). 
In this case, and  to be able to calculate the radial dependence  of  successive transfer, everything has to be translated 
in terms of single-particle motion and associated absolute separation energies and radial wave functions in systems with different 
relative coordinates.

 If the $k-$mass  connected with the 
 Perey-Buck energy-dependent term \cite{Mahaux}  already made the concept of a single mean field potential  
somewhat illusory (App. E), consider
the difficulties  one is  confronted with in attempting at translating  into a single-particle motion description 
 inside  a common potential, 
independent motion of Cooper pairs, composite bosonic particles with binding energies of the order  
 of one tenth of the Fermi energy
\footnote{Within this context we note that in $^{120}$Sn
the two-neutron separation energy is $S_{2n} =$ 15.6 MeV, while $S_{1n}=$ 9.1 MeV, i.e. ($2\times S_{1n}) - S_{2n}=$ 2.6 MeV.} 
 ($\approx 2 \Delta/\epsilon_F \approx$ 3~MeV/36 MeV)
and a correlation length of tens of fm, 
subject to a strong external field of radius $R_0 \approx $ 6 fm and  depth  $ \approx 50$ MeV.
 A way out  to this situation is provided by the fact  that in superfluid nuclei, one is not very far  from an independent particle picture.
As a consequence, no major errors are introduced in treating the system accordingly. 
Also in keeping with the fact that transfer  takes place through the single-particle field \cite{Poteletal2013}.

Summing up, while one does not know how to calculate the mass of the nucleus,  
one can accurately calculate   $U_j(118) V_j(120)$, as well as the relative value 
of the clothed  single-particle energies. In keeping with the fact that renormalised NFT which makes 
use of NG equation
correctly reproduces the quasiparticle energies, the Fermi 
energy of the single-particle potential used to generate  the radial wave function is adjusted 
so that the least bound state has the experimental separation energy $S_n$. 
 Within the unified picture of structure and reactions (NFT (r+s), \cite{Physica_scripta}), dressing f the radial wavefunctions give rise to the correct formfactors for transfer processes. While these effects are small for $^{120}$Sn, there are overwhelming in other situations, e.g. that of halo nuclei \cite{11Be}.


\section{Cooper pair population of pairing rotational bands: BCS,HFB and NG}\label{Section6}

In what follows we analyse the stability of the order parameter as probed by Cooper pair transfer.

\subsection{BCS}
Starting from 
a HF calculation with the  SLy4  interaction (Table \ref{table_novert2}, second column) we solve the BCS equations, and thus determine the corresponding occupation numbers $U_a(G)$ and $V_a(G)$ (Table \ref{table_novert2}, last two columns) with a schematic 
monopole pairing force of strength $G = 0.26 $ MeV, adjusted to fit the empirical three-point value $\Delta^{exp} \approx 1.4 $ MeV.
\subsection{HFB}
Making use of the same Skyrme interaction and of the $v_{14}$ Argonne, $^1S_0$ NN-potential and neglecting the influence of the bare
pairing force in the mean field,
the HFB equation was solved. As a result, this step corresponds to an extended BCS calculation over the HF basis, allowing for the 
interference between states of equal quantum numbers $a (\equiv lj)$, but different number of nodes $(k,k')$. We include $(N_a)$ states
(for each $a$) up to $\approx $1 GeV, to properly take into account  the repulsive core of $v_{14}$ and be able to accurately calculate 
$\Delta^{HFB}$. As a consequence, one obtains a set of quasiparticle energies $E^{\mu}_a$, with the quasiparticle index $(\mu  = 1,2... N_a)$.
To each quasiparticle $\alpha^+_{a,\mu} = \sum_{k=1}^{N_a} (U_a^{\mu,k} a^+_{a,k} - V_a^{\mu,k} a_{a,k})$ is associated an array
of quasiparticle amplitudes $U_a^{\mu,k}$ and $V_a^{\mu,k}$ which are the components of the quasiparticles over the HF
basis states $\phi^a_k = \langle \vec r | a^+_{a,k} |0\rangle   (\equiv \langle \vec r |a,k\rangle  ).$ Going to the canonical basis, where the density 
matrix takes a diagonal form, we look for the state having the largest value of the abnormal density, $(UV)_{max}$.
As a rule, for a well-bound nucleus such as $^{120}$Sn, this canonical state is  the quasiparticle state having 
the lowest value of the quasiparticle energy. 
The label $k$ then drops  because there is only one orbital for a given value of $a( \equiv  (lj))$. This implies that  
 the bare quasiparticle amplitudes can be characterised  simply by 
 $U_a,V_a$ and the associated state dependent value of the 
bare pairing gap is equal    $\Delta^{bare}_a = 2 U_a V_a E_a$.  
The values of $(E_a)_{min}$ and $(V_a(v_{14}))_{max}$ 
for the five valence orbitals are reported in Table \ref{table_novert2}. 
 
\subsection{Renormalized NFT and NG}

We now go beyond mean field and include the particle-vibration coupling leading to retardation phenomena 
both in self energy as well as in induced interaction processes. The vibrational modes are calculated in QRPA making
use of empirical (WS) single--particle levels, BCS with constant $G$ and multipole--multipole separable interactions of essentially
self--consistent strength \cite{Bohr1975} which reproduce the observed properties of the low--lying collective states.

To be able to treat the variety of possible situations we return to the full HFB basis. In this basis the $\omega-$dependent self energy 
has the following matrix structure \cite{Idini_tesi},
\begin{equation}
\hat{\Sigma}^a_{\mu,\mu'}(\omega) =  \left( \begin{array}{c c} 
 \Sigma^{11,a}_{\mu,\mu'}(\omega) &  \Sigma^{12,a}_{\mu,\mu'}(\omega) \\
 \Sigma^{21,a}_{\mu,\mu'}(\omega) &  \Sigma^{22,a}_{\mu,\mu'}(\omega)
\end{array} \right),
\end{equation}
the Dyson equation,
\begin{equation}
 \hat{G}^a(\omega+i\eta)= \left[\omega+i\eta - \hat{H}_{HFB} - \hat{\Sigma}^a(\omega+i\eta)\right]^{-1},
 \label{Green}
\end{equation}
providing the connection to the  corresponding Green's function matrix.
The imaginary part of this function is related to the strength functions 
that define energies and weights of the dressed quasiparticles,
\begin{eqnarray}
\tilde S^{a,+}_{k,k'}(\omega) = -\frac{\Im m}{\pi}\{ \sum_{\mu,\mu'} & G^{11,a}_{\mu,\mu'} U_{a}^{\mu,k}U_{a}^{\mu',k'} 
-  G^{12,a}_{\mu,\mu'} U_{a}^{\mu,k}V_{a}^{\mu',k'} \nonumber \\ & -  G^{21,a}_{\mu,\mu'} V_{a}^{\mu,a}U_{a}^{\mu',a'} 
+ G^{22,a}_{\mu,\mu'} V_{a}^{\mu,a} V_{a}^{\mu',k'} \},
\label{eq:S+} \\
\tilde S^{a,-}_{k,k'}(\omega) = -\frac{\Im m}{\pi}\{ \sum_{\mu,\mu'} & G^{11}_{\mu,\mu'} V^{\mu,k}_{a}V^{\mu',k}_{a'}
 +  G^{12}_{\mu,\mu'} V^{\mu,k}_{a}U^{\mu',k}_{a'} \nonumber \\ & +  
 G^{21}_{\mu,\mu'} U^{\mu,k}_{a}V^{\mu',k}_{a'} + G^{22}_{\mu,\mu'} U^{\mu,k}_{a}U^{\mu',k}_{a'} \}, 
 \label{eq:S-} \\
\widetilde{S}^a_{k,k'}(\omega) = -\frac{\Im m}{\pi}\{ \sum_{\mu,\mu'} & G^{11}_{\mu,\mu'} u^{\mu,k}_{a}v^{\mu',k}_{a'} 
+  G^{12}_{\mu,\mu'} U^{\mu,k}_{a}U^{\mu',k}_{a'} \nonumber \\ & -  
G^{21}_{\mu,\mu'} V^{\mu,k}_{a}V^{\mu',k}_{a'} - G^{22}_{\mu,\mu'} V^{\mu,k}_{a}U^{\mu',k}_{a'} \},
\label {eq:S}
\end{eqnarray}
where $\tilde S^{a,+}_{k,k'}(\omega)$, $\tilde S^{a,-}_{k,k'}(\omega)$ and $\widetilde S^a_{k,k'}(\omega)$ 
 play the role of  the probability density of the dressed quasiparticle, quasihole and of the corresponding 
 anomalous 
 component. It is also possible to express $\hat{\Sigma}$ as a function of $\tilde S^+, \tilde S^-$ and $\tilde S$ \cite{Idini_tesi}.
Thus one can carry out an iterative, self-consistent  procedure to calculate quasiparticle  renormalization, 
accounting for the so called rainbow series. 
This formalism does not assume the validity of the quasiparticle approximation, and iterates the solutions of the Dyson equations on the ansatz of continuous strength functions. 
However, close to the Fermi energy, quasiparticles peaks in the strength functions are clearly identifiable  due to their  characteristic Lorentzian shape,
as implied by the extension to the complex plane introduced in  (\ref{Green})
in  terms of the  parameter $\eta$ \cite{Fetter}.
Fitting  these peaks, one can determine the centroid energy $\tilde{E}_{a(n)}$
(dressed quantities labeled with a tilde carry a sum over $\mu$-values (see. Eq. (\ref{eq:S+})-(\ref{eq:S}))
 and associated 
width $\tilde{\Gamma}_{a(n)}$ for the fragment $n$, 
as well as its occupation amplitudes $\tilde{u}^{k}_{a(n)}$ and $\tilde{v}^{k}_{a(n)}$.

Alternatively, one can obtain the same result, still with an accuracy fixed by the $\eta-$parameter, but this time in terms of individual
levels
solving 
(at the last iteration) the eigenvalue Nambu-Gorkov problem,
\begin{equation}
\left( \hat{H}_{HFB} + \hat \Sigma^a(\widetilde{E}_{a(n)}) \right)_{k,k'} \left(\begin{array}{c} x^{k'}_{a(n)} \\ y^{k'}_{a(n)} 
\end{array}\right) = \widetilde{E}_{a(n)} 
\left(\begin{array}{c} x^{k}_{a(n)} \\ y^{k}_{a(n)} \end{array} \right).
\label{eq:Nambu}
\end{equation}

\begin{figure}[h!]
\centerline{\includegraphics[width=0.6\textwidth]{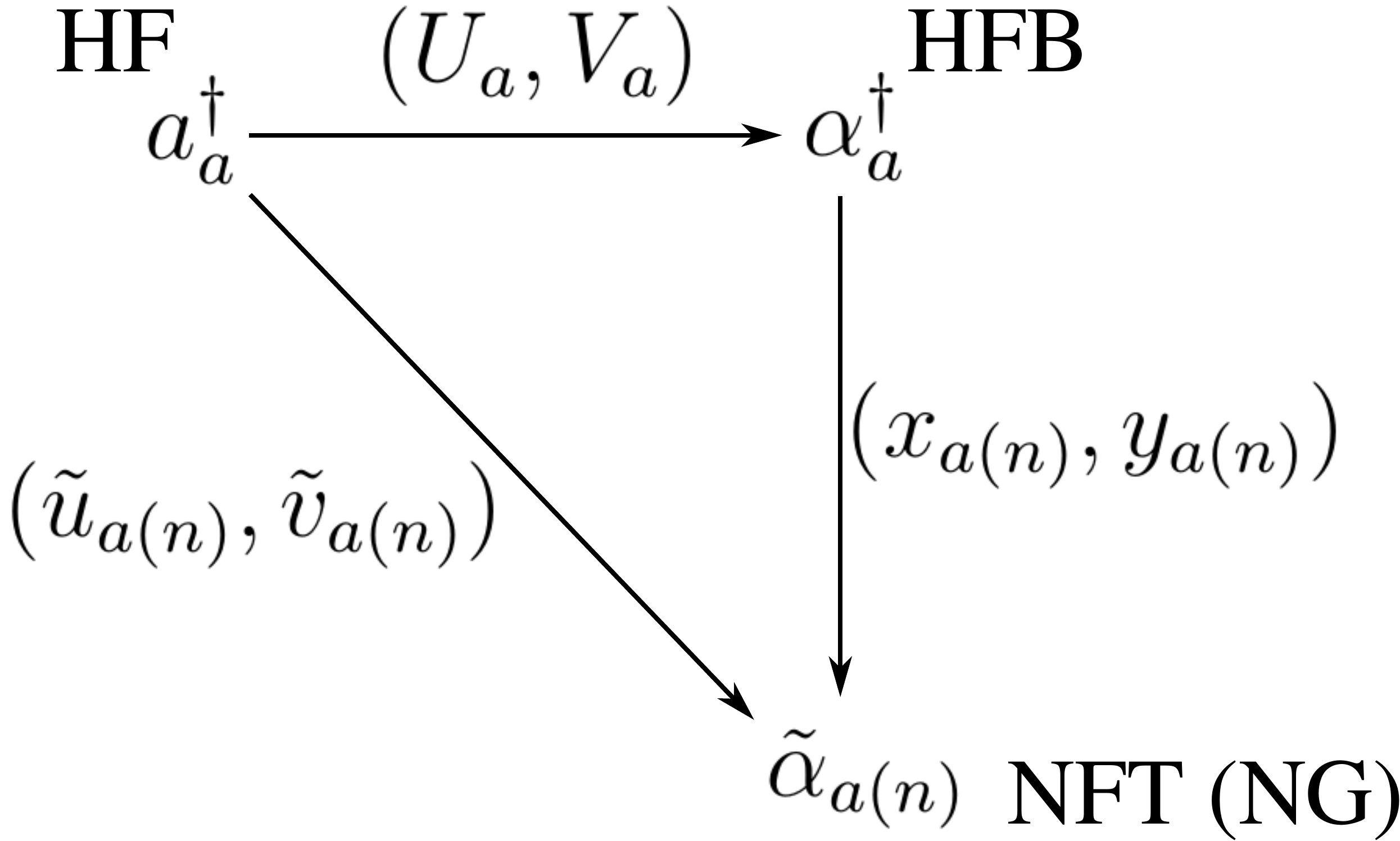}}
\caption{Schematic representation of the generalized quasiparticle transformations from independent 
particle states $a^+_a |0\rangle   = |a\rangle  $, to many-body clothed quasiparticle states $|\widetilde{a(\equiv (lj))(n)\rangle  } = \tilde \alpha^+_{a(n)} |\tilde 0\rangle  $,
($ \tilde \alpha^+_{a(n)} = \tilde u_{a(n)} a^+_a - \tilde v_{a(n)} a_a)$, made also in terms of a two-step protocol used in the present paper,
implemented in terms of a quasiparticle transformation from Hartree-Fock  ($a_a^+$) to Hartree-Fock-Bogoliubov 
($\alpha_a^+$) and a (self energy based) rotation (see Eq. (\ref{eq:transform})). } 
\label{fig:transform}
\end{figure}

The above formalism provides a most general framework to deal with the nuclear many-body problem, also in situations 
in which repulsive core and $\omega-$dependent soft modes mediated interactions are both active (see e.g. \cite{Pastore:08}). In the case of well bound nuclei lying along the stability valley as in the present case, 
the above equations can be simplifies turning to the canonical basis and, in keeping with the fact that the particle-vibration couplings
are mostly effective in a small region around the Fermi energy, restricting to the valence orbitals. 
Within this scenario, 
we  introduce the  shorthand notation $\Sigma^{ij}_{a(n)}  \equiv {\Sigma^{ij,a}(\tilde E_{a(n)})}$ for $i,j=1,2$  and note that is 
convenient to define the renormalised
quasiparticle amplitudes associated with a given solution $a(n)$,  as 
\begin{eqnarray}
\tilde u_{a(n)} = x_{a(n)} U_a - y_{a(n)} V_a, \nonumber \\
\tilde v_{a(n)} = x_{a(n)} V_a + y_{a(n)} U_a.
\label{eq:transform}
\end{eqnarray}
The above quantities are the quasiparticle amplitudes of the renormalised state $|\tilde a(n)\rangle  $.
The total quasiparticle strength associated with the $n-th$ fragment is (see Fig. \ref{fig:transform}) 
\begin{equation}
 \tilde N_{a(n)} = \tilde u^2_{a(n)}  + \tilde v^2_{a(n)} .
\end{equation}

The matrix elements of the total self energy rotated into the canonical basis  and identified in terms of primed quantities 
 including the bare interaction and the particle-phonon coupling, are given by 
\begin{eqnarray}
\tilde \Sigma^{11 '}_{a(n)}  = U_a^2 \tilde \Sigma^{11}_{a(n)} + V_a^2 \tilde \Sigma^{22}_{a(n)} -2 U_a V_a \tilde \Sigma^{12}_{a(n)}, \nonumber \\
\tilde \Sigma^{22 '}_{a(n)}  = U_a^2 \tilde \Sigma^{22}_{a(n)} + V_a^2 \tilde \Sigma^{11}_{a(n)} +2 U_a V_a \tilde \Sigma^{12}_{a(n)}, \nonumber \\
\tilde \Sigma^{12 '} _{a(n)}  = \Delta^{bare}_a +(\tilde \Sigma^{12 }_{a(n)})'_{ind},
 \end{eqnarray}
where we have defined
\begin{equation}
(\tilde \Sigma^{12}_{a(n)})'_{ind}   \equiv \tilde \Sigma^{12}_{a(n)} (U^2_a - V^2_a) + U_a V_a ( \tilde \Sigma^{11}_{a(n)} - \tilde \Sigma^{22}_{a(n)} ).
\end{equation}
The total pairing gap is   equal to 
\begin{equation}
\tilde \Delta_{a(n)}' = \tilde Z_{a(n)}  \tilde \Sigma^{12 '}_{a(n)},    
\end{equation}  
 the $Z$-factor \cite{Terasakietal2002} being
\begin{equation}
\tilde Z_{a(n)} =
\left( 1 - \frac{\tilde \Sigma^{odd}_{a(n)}} {\tilde E_{a(n)}} \right)^{-1},
\end{equation}
where 
\begin{equation}
\tilde \Sigma^{odd}_{a(n)} = \frac{\tilde \Sigma^{11}_{a(n)} + \tilde \Sigma^{22}_{a(n)}}{2}.
\end{equation}
It is of notice that for levels close to the Fermi energy $\Sigma^{odd}/\tilde E_{a(n)}$
approaches a derivative, and the physical role of $Z_{a(n)}$ approaches that of $N_{a(n)}$,
namely the quasiparticle component in the many-body renormalized quasiparticle state $ |\tilde a(n)\rangle  $.

We can identify two contributions to the pairing gap $\tilde \Delta_{a(n)}' $:
\begin{equation} 
\tilde \Delta_{a(n)}'  = [\tilde Z_{a(n)}  \Delta^{bare}_a ] +  [\tilde  Z_{a(n)} (\tilde \Sigma^{12 '}_{a(n)})_{ind}].
\end{equation} 
The first one  is related to the pairing gap associated with the  bare force and
quenched by the many--body effects which cloth the bare interacting nucleons. 
The second contribution obeys a generalised gap equation \cite{Idinietal2012},
\begin{equation}
(\tilde \Sigma^{12 '}_{a(n)})_{ind} = - \sum_{b,m} \frac{(2j_b+1)}{2} 
\langle b(m) {\overline{b(m)}} | v_{ind} | a(n) \overline{ a( n)}\rangle  
 \tilde u_{b(m)} \tilde v_{b(m)},
\end{equation}

where the induced interaction $v_{ind}$ is associated with the exchange of collective vibrations between 
pairs of nucleons moving in time reversal states. It can be written as (see App. G), 
\begin{align}\label{eq28}
&\langle b(m) \overline{b(m)} | v_{ind} | a(n) \overline{a(n)}\rangle  
 \nonumber \\
&=\sum_{\lambda,\nu} \frac{2 |h(a,b\lambda \nu)|^2} {(2j_b+1)} 
\left[ \frac{1} {\tilde E_{a(n)}  - \tilde E_{b(m)}  - \hbar \omega_{\lambda \nu}} - 
\frac{1} {\tilde E_{a(n)}  + \tilde E_{b(m)}  + \hbar \omega_{\lambda \nu}}  \right],
\end{align}
where $h(a,b\lambda \nu)$ denotes the matrix element coupling  the particle $a$
to the configuration $(b \otimes \lambda \nu)_a$, while the energy of the $\nu$-th phonon of multipolarity $\lambda$ 
is denoted $\hbar \omega_{\lambda \nu}$ \cite{Bohr1975}.  Concerning vertex correction to both $v_{ind}$ and $v_{bare}$
we refer to Appendix H.

The selection of the basis $|\tilde a(n)\rangle   = \tilde \alpha^+_{a(n)}|\tilde 0\rangle  $ through 
the rotation (\ref{eq:transform}) allows the eigenvalues of (\ref{eq:Nambu}) to retain the standard BCS relation,
namely
\begin{equation}
\tilde{E}_{a(n)} = \sqrt{ (\tilde \epsilon_{a(n)}' - \epsilon_F) ^2 +  (\tilde\Delta_{a(n)}')^2 },
\end{equation} 
the renormalised quasiparticle energy being
\begin{equation}
\tilde \epsilon_{a(n)}'  - \epsilon_F = \tilde Z_{a(n)} 
\left[( \epsilon_a - \epsilon_F) + \tilde \Sigma^{even'}_{a(n)}  \right],
\end{equation}
where 
\begin{equation} 
\Sigma_{a(n)}^{even'} = \frac{\tilde \Sigma^{11'}_{a(n)}  - \tilde \Sigma^{22'}_{a(n)}}{2}.
\end{equation} 
It is of notice that $\tilde \Sigma_{odd}$ is invariant under the rotation (\ref{eq:transform}), the same being
true for $\tilde Z_{a(n)}$ and $\tilde E_{a(n)}$, while this does not apply to $\Sigma_{a(n)} ^{even '} $.   


\begin{table}[h!]
\begin{center}
\begin{adjustbox}{max width=\textwidth}
\begin{tabular}{|c  | c | c | c | c |  c | c |  c | c | c | c |c |c |c |c|c|}
\hline 
$a$ &$\epsilon_{a}$&$\tilde \epsilon_{a(n)}$ & $n$ & $\tilde E_{a(n)}$ &   $\tilde u^2_{a(n)}$  &  $\tilde v^2_{a(n)}$  & $N_{a(n)}$& $Z_{a(n)}$&  $\tilde \Delta_{a(n)} $ & $E_{a}  ({v}_{14})$&  $U^2_{a} ({v}_{14}) $   
   &$V^2_{a}({v}_{14}) $  &$E_{a}$(G)&$U^2_{a}$(G) & $V^2_{a}$(G) \\
\hline
$d_{5/2} $ &-10.7& -9.4 &  1 & $2.55$  & 0.06 & $ 0.28$& 0.34& 0.60& 1.96 &3.12 &  0.03 & 0.97 & 3.09 & 0.06 & 0.94 \\ \cline{2-14} 
&  &-9.9 &  2 & $2.75$  & 0.01 & $ 0.10$&0.11& &1.80 && & & & &\\ \cline{2-10}
&  &-10.5 & 3 & $3.19$  & 0.01 & $ 0.10$& 0.11& & 1.68 && & & & & \\ \cline{2-10}
&  & -10.6 & 4 & $3.36$  & 0.01 & $ 0.07$& 0.08&  & 1.88 & & & & & &\\ \cline{2-10} 
 & &-11.2 &  5 & $3.95$  & 0.01 & $ 0.07$ & 0.08&  &  1.97 & & & & & &  \\ \cline{2-10}
  & &-12.4&  6& $4.77$  & 0.0 & $ 0.07$ & 0.07&  & -1.29 & & & & & &  \\ \cline{2-10}
    & &-12.7 & 7& $4.98$  & 0.0 & $ 0.09$&0.09&  &  -0.61 & & & & & &    \\ \hline \hline
$g_{7/2} $ &  -10.1 & -9.3 & 1 & $2.10$  & 0.09 & 0.59& $ 0.68$& 0.78& 1.43  & 2.56 & 0.06  & 0.94 &2.54 &  0.09  & 0.91 \\ \cline{2-14} 
& &-10.6 & 2 & $2.83$  & 0.00 & $ 0.08$ & 0.08 &  &0.34  &  & & & & & \\ \cline{2-10}
& & -9.9& 3 & $3.20$  & 0.00 & $ 0.0 $  &0.0& & -2.40  & & & & & \\ \cline{2-10}
& & -11.2& 4 & $3.50$  & 0.00 &  0.11 & 0.11  & & 0.97 &  & & & & & \\  \hline \hline
$s_{1/2}$ & -9.0  & -8.4 & 1 & $1.80$  & 0.26 & $ 0.53$ &0.79& 0.72& 1.69 &1.61 &  0.13  & 0.87 & 1.79 & 0.22  & 0.78 \\ \cline{2-14}
& & -10.4& 2 & $2.84$  & 0.00 & $ 0.04$& 0.04& & -1.03  & & & & & & \\ \cline{2-10}
& &-10.1 & 3 & $3.20$  & 0.00 & $ 0.0$ &  0.0& & -2.20 & && & & & \\ \cline{2-10}
& &-12.4 & 4 & $4.64$  & 0.00 & $ 0.07$ & 0.07 & & -0.46  & && & & & \\ \hline \hline
$d_{3/2} $ & -8.5 &-7.9  &  1 &$1.48$  & 0.38 & $ 0.46$ & 0.84& 0.76& 1.48& 1.37 & 0.24 & 0.76 &1.57 &  0.33 & 0.67\\ \cline{2-14}
& &-7.5 & 2 &$2.75$  & 0.0 & $ 0.00$ & 0.0 & &-2.73  & & & &&& \\ \cline{2-10}
&  &-8.8 & 3 &$3.06$  & 0.0 & $ 0.01$ & 0.01&&  -2.88 & & & &&&\\ \cline{2-10}
& &-11.3 & 4 &$3.49$  & 0.0 & $ 0.05$ & 0.05&&  -0.14 & & & &&&  \\ \hline \hline
 $h_{11/2} $   & -7.1& -7.2 & 1 & $1.64$ & 0.57 & $ 0.26$& 0.83& 0.79& 1.52   &1.34  &0.79&0.21 & 1.74 & 0.77 & 0.23 \\ \cline{2-14}
&  &-4.7 & 2 & $3.08$  & 0.09 & $ 0.00$ & 0.09 & &0.08&   & & &  & & \\ \cline{2-10}
&  &-9.6 & 3 & $3.97$  & 0.00 & $ 0.00$ &  0.0& &3.54 & & & & &&\\ \hline
\end{tabular}
\end{adjustbox}
\caption{\protect In the first four columns we list the orbital,  its HF  energy $\epsilon_{a}$ calculated with the SLy4 interaction, 
the renormalised energy  $\tilde \epsilon_{a(n)}$  of the main n-peaks resulting from the breaking of the strength due 
to renormalisation  effects.
   In the next six columns we list the renormalised  quasiparticle energies,  occupation factors, state dependent gap 
   of the lowest peaks associated with 
each of the five valence levels in $^{120}$Sn, carrying more than 5\% of the single-particle strength.
In the other columns we list quasiparticle energies and  occupation factors obtained in a HFB calculation with the Argonne interaction ($v_{14}$) 
($\Delta^{HFB} = $1.08 MeV)
and with a monopole force of  strength
 $G =$ 0.26 MeV, fitted to reproduce the empirical three point value $\Delta^{exp} \approx 1.45 $ MeV.
This last calculation is equivalent to BCS.
In all cases, the energy of the $d_{5/2}$ level has been shifted  by 600 keV towards the Fermi energy. 
In the renormalised calculation, spin modes have been
effectively taken into account by including  a repulsive monopole interaction of strength $G= 0.03 $ MeV acting on the valence orbitals 
in the solution of the Nambu-Gorkov equation 
(quantitative effect of spin modes \cite{Idini2015}).}
\label{table_novert2}
\end{center}
\end{table}

\begin{table}[h!]
\begin{center}
\begin{tabular}{|c  | c | c | c |c|}
\hline
& \multicolumn{4}{c}{B(a(1)) (($\alpha_0)_a$)}   \\
\hline
$a \equiv \{l_j\}$ & NFT(NG) & HFB($v_{14}) $& BCS(G)  & Z BCS(G)\\ 
\hline
$d_{5/2} $ &0.22 (0.39)& 0.29 (0.51) & 0.41 (0.71) &0.25(0.43)\\ \hline
$g_{7/2}$ & 0.46 (0.92) & 0.47 (0.95) & 0.57 (1.14)& 0.45(0.89)\\ \hline
 $s_{1/2}$ & 0.37 (0.37) & 0.34 (0.34)& 0.41 (0.41)& 0.30(0.30)\\ \hline
$ d_{3/2}$ & 0.59 (0.84) & 0.60 (0.85) & 0.66 (0.94)& 0.50 (0.71) \\ \hline
 $h_{11/2}$ & 0.95 (2.34) & 1.0 (2.44)  & 1.03 (2.52)&0.81(1.99) \\\hline
$\alpha_0$ & 4.83 (14) & 5.09(15) & 5.74(16) & 4.32(12) \\ \hline
\end{tabular}
\caption{ Two-nucleon spectroscopic amplitudes (Eqs. (\ref{eq:Bj}) and (\ref{eq:Btilde}) and contribution
to $\alpha_0 $ ($(2j_a +1/2)^{1/2} B(a)$) calculated making use of the quantities given in Table \ref{table_novert2}.
In the last row the value of $\alpha_0$ is reported while the percentage of the number 
of neutrons (i.e. 2$\alpha_0$/70) participating in the condensate is given in parenthesis.
In the last column the quantities worked out making use of the approximation (\ref{eq:uv}) for $\tilde u_{a(n)}$ and
$\tilde v_{a(n)}$ are given.}\label{table3x}
\end{center}
\end{table}

\begin{table}[h!]
{\begin{tabular}{|c|c|c|c|c|c|c|c|c|c|c|c|c|}
\cline{2-13} 
\multicolumn{1}{c|}{}& \multicolumn{12}{|c|}{$^{A}$Sn($p,t)^{A-2}$Sn}           \\
\cline{2-13} 
\multicolumn{1}{c|}{} & $V$ & $W$ &  $V_{so}$ &  $W_d$ &  $r_1$ &  $a_1$ &  $r_2$ &  $a_2$ &  $r_3$ &  $a_3$ &  $r_4$ &  $a_4$            \\
\hline 
$p,\;^A$Sn$\,^{a)}$ & $50$ & $5$ &  $3$ &  $6$ &  $1.35$ &  $0.65$ &  $1.2$ &  $0.5$ &  $1.25$ &  $0.7$ &  $1.3$ &  $0.6$ \\
\hline 
$d,\;^{A-1}$Sn$\,^{b)}$ & $78.53$ & $12$ &  $3.62$ &  $10.5$ &  $1.1$ &  $0.6$ &  $1.3$ &  $0.5$ &  $0.97$ &  $0.9$ &  $1.3$ &  $0.61$ \\
\hline 
$t,\;^{A-2}$Sn$\,^{a)}$ & $176$ & $20$ &  $8$ &  $8$ &  $1.14$ &  $0.6$ &  $1.3$ &  $0.5$ &  $1.1$ &  $0.8$ &  $1.3$ &  $0.6$ \\
\hline
  \end{tabular}}
    \captionsetup{singlelinecheck=off,justification=raggedright}
   \caption{\protect Optical potentials used in the calculation of the absolute  two--nucleon transfer differential cross sections. 
   The quantities $V,W,V_{SO},W_d$ are in MeV while the remaining quantities are in fm.
   The nuclear term of the optical potential was chosen to have the form
   $U(r)=-Vf_1(r)-iWf_2(r)-4iW_d\;g_3(r)-\left(\frac{\hbar}{m_\pi c}\right)^2 V_{so}\frac{g_4(r)}{a_4r}\mathbf{l\cdot s},$
   with $f_i(r)=\frac{1}{1+e^{(r-R_i)/a_i}};\quad g_i(r)=\frac{e^{(r-R_i)/a_i}}{\left(1+e^{(r-R_i)/a_i}\right)^2},$
   and $m_\pi$ the pion mass, while $R_i=r_iA^{1/3}$, $A$ being 
   the mass number of the heavy nucleus in the corresponding channel. 
   The Coulomb term is taken to be the electrostatic potential generated by an uniformly charged sphere of radius $R_1$.
   $a)$ ref. \cite{Guazzoni} $b)$ ref.\cite{An}.} 
\label{table4}
\end{table}

\begin{figure}[h!]
\includegraphics[width=0.8\textwidth]{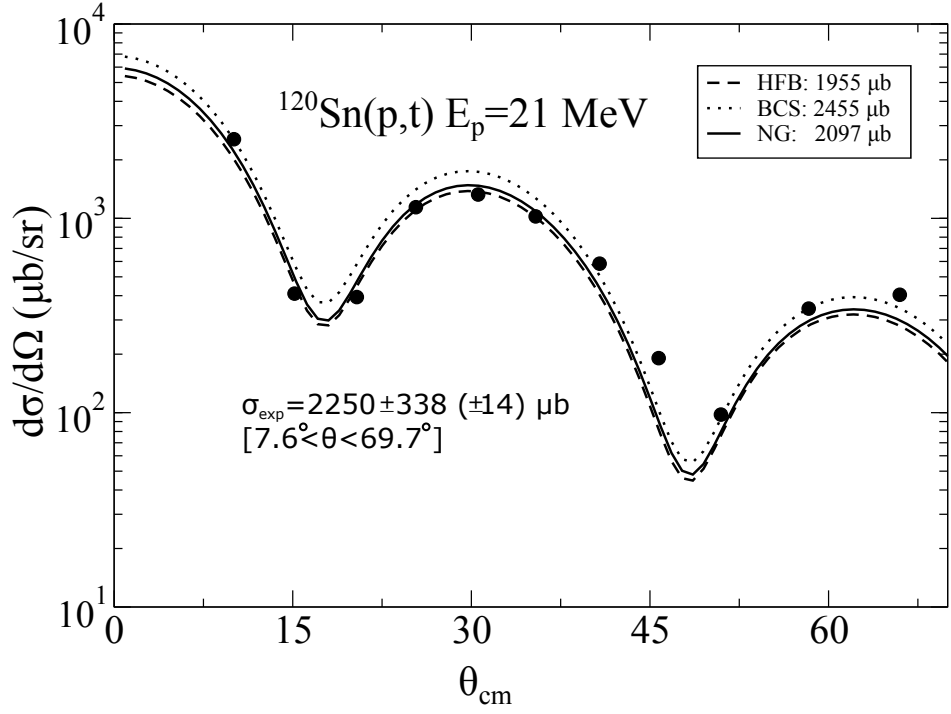}
\caption{Absolute differential  cross sections associated with the reaction
$^{120}$Sn(p,t)$^{118}$Sn(gs) calculated making use of the BCS, HFB and renormalised NFT(NG) spectroscopic 
amplitudes (Table \ref{table3x}) and global optical parameters (Table \ref{table4}), in comparison with the experimental findings (solid dots) \cite{Guazzoni}.}
\label{fig:cross}
\end{figure}

\begin{figure}[h!]
\centerline{\includegraphics[width=0.7\textwidth]{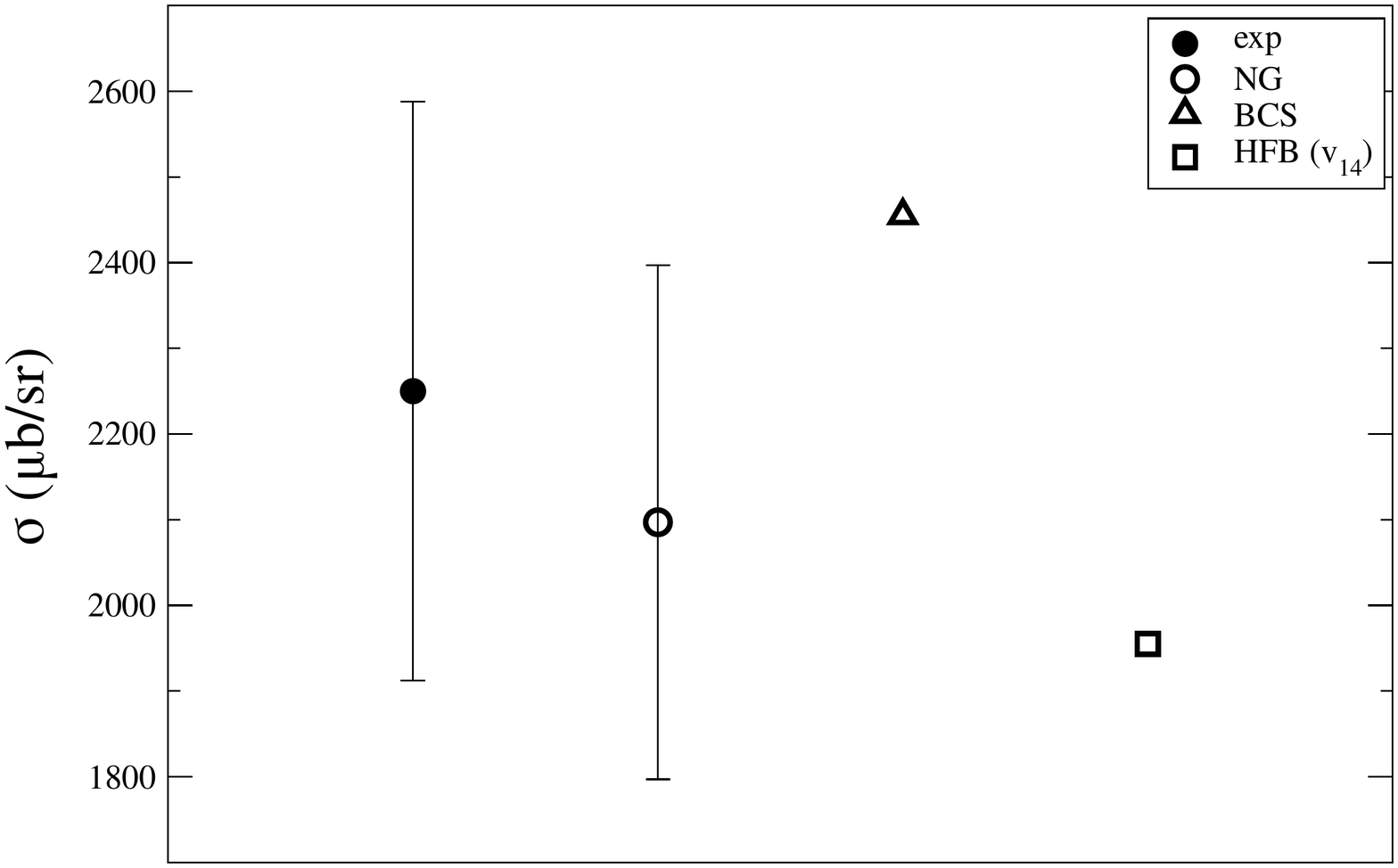}}
\caption{Integrated absolute  cross sections associated with the reaction
$^{120}$Sn(p,t)$^{118}$Sn(gs) (see caption to Fig. \ref{fig:cross}). The error ascribed  to
the NFT(NG) theoretical results stems from  the uncertainties in the calculation of the 
two-neutron transfer  spectroscopic amplitudes estimated from the 
variations the contribution of  spin modes associated with  different  Skyrme interactions induce in the 
$B-$coefficients.}\label{fig5}
\end{figure}

The results obtained from the solution  of the Nambu-Gor'kov equation are collected in Table \ref{table_novert2}, together 
with those if HFB and BCS.
 The fragments carrying the largest fraction of the quasiparticle strength 
associated with each of the five valence orbitals of unperturbed energy $\epsilon_a$  are listed in order 
of increasing energy.
For each fragment the value of the   renormalised quasiparticle energy  $\tilde E_{a(n)}$ and of 
the renormalized quasiparticle amplitudes $\tilde u_{a(n)}, \tilde v_{a(n)}$ are  provided, together  with those of the  renormalised single-particle energy $\tilde \epsilon_{a(n)}'$ and
 of the renormalised pairing gap $\tilde \Delta_{a(n)}'$.

The formalism outlined above has been used to compute 
the two--nucleon transfer spectroscopic amplitudes   
\begin{equation}
 \tilde{B}(a(n)) = \sqrt{\frac{2j_a + 1}{2} } \tilde{u}_{a(n)}\tilde{v}_{a(n)} = \sqrt{\frac{2j_a + 1}{2} } \int^{\tilde E_{a(n)}+\tilde \Gamma_{a(n)}/2}_{\tilde E_{a(n)}- \tilde \Gamma_{a(n)}/2} \widetilde{S}_a (\omega ) \textrm{d}\omega,
 \label{eq:Btilde}
\end{equation}
associated with the reaction $^{120}$Sn$(p,t)^{118}$Sn$(gs)$
between two members of the Sn--ground state pairing rotational band. 
The corresponding results are shown in Table \ref{table_novert2}, in comparison to those corresponding to the HFB
and BCS calcualtion. Making use of global optical potentials (Table \ref{table4}), the 
absolute differential 
cross sections were  calculated and are compared with the experimental findings in Fig. \ref{fig:cross}.
Theory reproduces the experimental findings essentially 
at the 10\% level (BCS 9.1\%, HFB 13\%, NFT (NG) 7\%), well within experimental errors (see  also Fig. \ref{fig5}).
The stability of the theoretical results is apparent.

\clearpage

\section{Discussion}

The spectroscopic results reported in Table \ref{table_novert2} testify to 
the important effects renormalisation of the single-particle 
states and of the pairing interaction have at the level of quasiparticles.
In spite of this, all three approaches (NFT(NG),HFB, BCS), notwithstanding 
their large differences in terms of many-body facets, predict essentially equally correct absolute two-nucleon
transfer cross sections, as testified by the results displayed in Figs.  \ref{fig:cross} and \ref{fig5}, where theory is
compared to experiment, 

It seems then fair to conclude that the quantity which controls the   specific excitation
of pairing rotational bands, namely the order parameter $\alpha_0$, in the sense of Cooper pair transfer amplitude (Sect. \ref{s3}), is essentially invariant, 
whether calculated within the framework of the simplest  one-pole quasiparticle (BCS) approximation, or
taking into account the variety of many-body renormalisation effects.

The emergence of a physical sum--rule is apparent (within this context see \cite{Broglia:72}, while for exact sum rules see \cite{Bayman:72,Lanford:77}).
Let us elaborate on this point.

Approximating
\begin{equation}
\tilde u_{a(n)} = \sqrt{N_{a(n)}} U_a \quad ; \quad   
\tilde v_{a(n)} = \sqrt{N_{a(n)}} V_a,
\label{eq:uv}
\end{equation}
and  
\begin{equation}
N_{a(n)} \approx Z_{a(n)} \approx Z_{\omega}, \quad (\epsilon_a \approx \epsilon_F)
\label{Na}
\end{equation}
one can write,
\begin{equation}
\alpha_0 = \sum_{a,n} \frac{2j_a+1}{2} \tilde u_{a(n)} \tilde v_{a(n)}  = \frac{N(0)}{Z_{\omega}}
\int d\epsilon \frac{2j_{\epsilon}+1}{2} \tilde u_{\epsilon} \tilde v_{\epsilon},\label{eq:alpha01}
\end{equation}
where $N(0)/Z_{\omega}$ is the effective density of levels at the Fermi energy \cite{Schuck}. With the help
of Eq. (\ref{Na}) one obtains,
\begin{equation}
\alpha_0 = \frac{N(0)}{Z_{\omega}} Z_{\omega}
\int d\epsilon \frac{2j_{\epsilon}+1}{2} U_{\epsilon} V_{\epsilon}
\approx \sum_a \frac{2j_a +1}{2} U_a V_a.
\label{eq:alpha0} 
\end{equation}
Using each  term of the expressions (\ref{eq:alpha01}) and (\ref{eq:alpha0}) as weighting factors of the corresponding two--nucleon transfer formfactors, in keeping with the unified structure--reaction physical interpretation of $\alpha_0$ (Sect. \ref{s3}), and that (see Figs. \ref{fig:cross} and \ref{fig5}) $|\sigma_i-\sigma_{exp}|/\sigma_{exp}$ is equal to 0.09, 0.13 and 0.07 ($i=$ BCS, HFB, NG), the relative errors of the associated two--nucleon transfer amplitudes $\alpha_0(\sim\sqrt{\sigma})$ are 4.5\%, 6.5\% and 3.5\%. Within this context, it is of notice that the fact that the HFB result lies closer to the NG one than BCS does, is a simple consequence of NG being based on HFB. 

Furthermore, because the matrix elements of $v_{14}$ for configurations based  on the valence orbitals is essentially state independent together with the fact that
$Z^2 \approx 0.5$, setting $v_{ind} = 0$, one expects for the renormalised (NFT(NG)) cross section a value 
$\approx 1000 \mu b$ (0.5 $\times \sigma_{HFB}$), precluding the above accuracy. Consequently, at the basis 
of the validity of (\ref{eq:alpha01})--(\ref{eq:alpha0}) and thus of the conservation of two--nucleon transfer  amplitudes in going 
from BCS mean field to NFT(NG) many-body, medium renormalization representations, one also finds the central role  played by the induced pairing
interaction. 
\clearpage

\setcounter{equation}{0}
\appendix 
\chapter{\large \bf  Appendix A. Off diagonal long range  order (ODLRO)}

\vspace{1cm} 

\makeatletter
\renewcommand{\theequation}{A\@arabic\c@equation}
\makeatother

\makeatletter
\renewcommand{\thefigure}{A\@arabic\c@figure}
\makeatother


The challenge solved by Schrieffer  \cite{Schrieffer1964} in his contribution to BCS was that of writing, starting from 
Cooper  single pair solution to pairing \cite{Cooper1956},  a many-particle wave function in which  each electron moving 
close to the Fermi energy participated in the condensate. 
The main problem
is that $N-$fixed many-body wave functions cannot have a definite phase. But if one uses a coherent 
state representation it is possible  to describe a condensate with a definite phase. 
Schrieffer found a way to
write down a coherent state of fermion pairs, namely 
(it is of notice that primed quantities are again being used , see footnote p. 3)
\begin{equation}
|BCS\rangle    =  \Pi_{\nu\rangle  0}  (U_{\nu}   +  V_{\nu}  a^+_{\nu}  a^+_{\bar \nu}) |0 \rangle  .
 \end{equation}
Introducing the phasing 
\begin{equation}
U_{\nu} = |U_{\nu}| =  U'_{\nu},     \quad V_{\nu} =e^{-2i\phi}   V'_{\nu} ( V'_{\nu}    \equiv |V_{\nu}|),
\end{equation}
one can write 
\begin{eqnarray}
|BCS(\phi)\rangle  _{\cal K}  =    \Pi_{\nu\rangle  0}  (U'_{\nu}  +  V'_{\nu}   e^{-2i\phi}  a^+_{\nu}  a^+_{\bar \nu}) |0 \rangle   
\\
=  \Pi_{\nu\rangle  0}  (U'_{\nu} +  V'_{\nu}  a'^+_{\nu}  a'^+_{\bar \nu}) |0 \rangle  
= |BCS(0)\rangle  _{\cal K'} , 
 \end{eqnarray}
where $\cal K$ and $\cal K'$ label the laboratory and the intrinsic (body-fixed BCS, deformed 
state in gauge space)
 frame of reference, while $a'^+_{\nu} = {\cal G} (\phi) a^+_{\nu}
{ \cal G}^{-1}  (\phi) = e^{-i \phi } $
$a^+_{\nu}  (a'^+_{\bar  \nu} $
$= e^{-i \phi }  $
$a^+_{\bar \nu}) $ is a creation operator  
 referred to this intrinsic frame. The operator $ {\cal G}  (\phi)$ = exp$({-i  {\hat N} \phi })$
 induces a rotation of angle $\phi$ in gauge (two-dimensional) space (gauge transformation), with $\hat N$ being 
 the number of particle operator. The states $|\nu\rangle  $ and $|\bar \nu\rangle  $, connected  
 by the time-reversal
 operator, have the same energy (Kramers' degeneracy). 
 
 A property of the above wave function, which has been given the name "off-diagonal long-range order" (ODLRO)  
 \cite{Yang1962}
 is of crucial importance regarding the physics at the basis of BCS condensation
 \footnote{Within this context, let us quote from Leon Cooper's contribution to  the volume {\bf BCS: 50 years} \cite{BCS50}:  
 "It has become 
 fashionable  ... to assert ... that once gauge symmetry is broken, the properties of superconductors follow ... 
 with no need to inquire into the mechanism by which the symmetry is broken. This is
 not ... true, since broken gauge symmetry might lead to molecule-like pairs 
 and a Bose-Einstein (BEC , Feshbach resonance see {\bf a)} below,  {\it our comment }) rather than BCS condensation 
 ... in 1957, we were aware that what is now called broken gauge symmetry would, under some 
 circumstances (an energy gap or an order parameter), lead to many of the qualitative
 features of superconductivity ... the major problem was to show how an energy gap, an order parameter
 of "condensation in momentum space" could come about .. to show... how the gauge-invariant
 symmetry of the Lagrangian could be spontaneously broken due to the interactions which were
 themselves gauge invariant". {\bf a)}  A Feshbach resonance is an enhancement in the scattering amplitude of a particle
 incident on   a target  - for instance, a nucleon scattering from a nucleus or an atom
 scattering form another one - when it has approximately the energy needed to create a quasi-bound
 state of the two-particle system. By making it feasible to precisely (Zeeman-tuned) control interactions, Feshbach resonances provide a tool for creating ultracold molecules and BECs.}. This property can be extracted from the BCS
 wave function in a number of ways (see e.g. \cite{Ange})
 

To introduce the subject, 
let us start by writing down operators  which create or annihilate pairs of fermions 
in the $\vec r$ representation, i.e. making use of \\
$\psi^+(\vec r)= \langle \vec r |a^+_{\nu} |0\rangle  $ 
and the Hermitian 
conjugate (see App. B). One can define the pair operator  (see Fig. A1)
\begin{equation}
P^+(\vec R) = \int d^3 r \phi(\vec r) \psi^+_{\nu} (\vec R + \vec r/2) \psi^+_{\bar \nu} (\vec R - \vec r/2),
\end{equation}
where $\phi(\vec r) $ is the pair wave function.
Thus $P^+ (\vec R)$ creates a spin singlet fermion pair where the particles are separated 
by the relative distance $\vec r$ and with centre of mass $\vec R$, i.e. 
\begin{equation}
\vec R = \frac{\vec r_1  + \vec r_2}{2} \quad , \quad  \vec r = \vec r_1 - \vec r_2,
\end{equation}
and thus 
\begin{equation}
\vec r_1 = \vec R + \frac {\vec r}{2} \quad , \quad r_2= \vec R - \frac {\vec r}{2}.
\end{equation}
One can now define a density matrix
\begin{equation}
\rho(\vec R - \vec R') = \langle P^+(\vec R) P(\vec R')\rangle  ,
\end{equation}
that is, a generalised particle (see Eq. (\ref{eqc1}), App. C)  density for pairs, the so called abnormal density, related to
the two-particle density
\begin{equation}
\rho_2 (\vec r_1 \sigma_1, \vec r_2 \sigma_2, \vec r_3 \sigma_3, \vec r_4 \sigma_4) =
\langle  \psi^+_{\nu} (\vec r_1) \psi^+_{\bar \nu} (\vec r_2) \psi^+_{\bar \nu'} (\vec r_3) \psi^+_{\nu'} (\vec r_4)\rangle  .
\end{equation}
Making use of Eq. (A.1) one obtains, 
\begin{eqnarray}
\rho(\vec R - \vec R') =  
\int d^3 r d^3 r\;' \phi(\vec r ) \phi(\vec r \;')  \nonumber \\
\times \rho_2(\vec R + \vec r/2, \sigma; \vec R - \vec r/2, -\sigma;
 \vec R\;' -  \vec r\;'/2, -\sigma; R\;'+ \vec r\;' /2, \sigma\;') .
\end{eqnarray}
The pair wave function $\phi(\vec r)$ vanishes when the relative distance becomes larger 
than the correlation length $\xi$. Thus, $\rho$ is different from zero provided that $r ( \equiv |\vec r_1 - \vec r_2|)$
and $r' ( \equiv |\vec r_1\;' - \vec r_2\;'|)$, 
 are smaller than $\xi$. But the pairs can be separated by any arbitrary distance. In other
words, ${\rm lim}_{|\vec R - \vec R'| \to \infty}  \rho(\vec R - \vec R\; ') \neq 0$, that is ODLRO. And this is what
the $|BCS\rangle  $ state ensures, in keeping with the fact that it describes independent pair motion, in which all pairs 
are in the same state, i.e.
\begin{eqnarray}
|BCS(\phi)\rangle  _{\cal K} = \left( \Pi_{\nu>0} U'_{\nu} \right) \{ 1 + ( \sum_{\nu>0}   
c'_{\nu} e^{-2 i \phi} a^+_{\nu} a^+_{\bar \nu} + \nonumber  \\
+\frac{1}{2!} (\sum_{\nu>0}  c'_{\nu} e^{-2 i \phi}  a^+_{\nu} a^+_{\bar \nu})^2 
+\frac{1}{3!} (\sum_{\nu>0}  c'_{\nu} e^{-2 i \phi}  a^+_{\nu} a^+_{\bar \nu})^3 + ...  \}
\label{BCSphi}
\end{eqnarray}
where
\begin{equation}
c'_{\nu} = \frac{V'_{\nu}} {U'_{\nu}}.    
\end{equation}
This is the mean field solution of  the pairing Hamiltonian. In other words,
the ground state of the mean-field pairing Hamiltonian
\begin{equation}
H_{MF}  = U + H_{11}, 
\end{equation}
where
\begin{equation}
U= 2 \sum_{\nu>0} (\epsilon_{\nu}  - \lambda) V^2_{\nu} - \frac{\Delta^2}{G} ,  
\end{equation}
and 
\begin{equation}
H_{11} = \sum_{\nu}  E_{\nu} \;  \alpha^+_{\nu}  \alpha_{\nu} .    
\end{equation}
Gauge symmetry restoration is obtained by taking into account the interaction
\begin{equation}
H''_p = \frac{G}{4} \left( \sum_{\nu>0} (\Gamma^+_{\nu}  - \Gamma_{\nu} ) \right)^2, 
\label{eqA12} 
\end{equation}
acting among the quasiparticles where $\Gamma^+_{\nu} = \alpha^+_{\nu} \alpha^+_{\bar \nu}$.
In fact, it can be shown that  
\begin{equation}
[H_{MF} + H''_p, \hat N ] = 0,
\label{eq:comm}
\end{equation}
$\hat N$ being the number of particle operator.
Diagonalising Eq. (\ref{eqA12})  in QRPA, i.e.
\begin{equation}
[H_{MF} + H''_p, \Gamma''^+_{n} ] = W''_n \Gamma^+_n,
\label{eqA13} 
\end{equation}
where
\begin{equation}
\Gamma''^+_n = \sum_{\nu>0} (a_{n \nu} \Gamma^+_{\nu} + b_{n \nu} \Gamma_{\nu}),
\end{equation}
the associated dispersion relation reads 
\begin{equation}
\sum_{\nu >0 } \frac{2 E_{\nu}}{(2E_{\nu})^2  - (W''_n)^2} = \frac{1}{G},
\label{EqA15}
\end{equation}
while  
\begin{equation}
a_{n \nu} = \frac{\Lambda''_n}{2E_{\nu} - W''_n} \quad, \quad 
b_{n \nu} = \frac{\Lambda''_n}{2E_{\nu} + W''_n},
\end{equation}
with
\begin{equation}
\Lambda''_n = \frac{1}{2} 
\left( \sum_{\nu>0} \frac{2E_{\nu} W''_n}{((2E_{\nu})^2 - (W''_n)^2)^2} \right)^{-1/2}.
\label{eq:dispers}
\end{equation}
The lowest - most "collective" - root of Eq. (\ref{EqA15}) has $W''_1=0$ (BCS gap equation),
the associated eigenstate being 
\begin{equation}
|1''\rangle   = \Gamma''^+_1 |0\rangle   = \Lambda''_1 \sum_{\nu >0} \frac{1}{2E_{\nu}} 
(\Gamma^+_{\nu} +\Gamma_{\nu}) |0''\rangle  .
\label{eqA17} 
\end{equation}
In keeping with the fact that, in QRPA, the number operator reads 
\begin{equation}
\tilde N = \Delta \sum_{\nu>0} \frac{1}{E_{\nu}} (\Gamma^+_{\nu} + \Gamma_{\nu}) +N_0,
\end{equation}
one can write 
\begin{equation}
|1''\rangle   = \Gamma''^+_1 |0\rangle  
= \frac{\Lambda''_1}{2 \Delta} (\tilde N - N_0) |0''\rangle  .
\label{eqA20} 
\end{equation}

A finite rotation in gauge space can be generated by a series of infinitesimal
operations induced  by the operator $  {\cal G} (\phi) = $ exp $( -i \hat N \phi)$,
i.e. 
\begin{equation}
{\cal G} (\delta \phi) \approx 1 - i {\hat N} \delta \phi . 
\end{equation}
Within this context, $i ({\cal G} (\delta \phi) - ( {\cal G} (0) -  i N_0 \delta\phi) ) = 
(\hat N - N_0) \delta \phi$,
where $\delta \phi = \Lambda''_1 /(2 \Delta) $ (Eq. (\ref{eqA20} ).
Because $\Lambda''_1$ diverges as $W''_1 \to 0$, $ (\hat{N}- N_0)|{\tilde 0 ''}\rangle   \approx
 (\tilde N -N_0) | {\tilde 0 ''}\rangle  \to 0$,
in keeping with Eq. (\ref{eq:comm}).    

Divergence in gauge angle implies that $\phi$ can have any value in the range 
$0-2\pi$. Consequently the system will be in a given member of a pairing rotational band,
e.g.

\begin{equation}
|\tilde N \rangle   \sim \int d\phi e^{i (N/2) \phi } |BCS(\phi)\rangle  _{\cal K} \sim O^{N/2} |0\rangle  ,   
\end{equation}
where 
\begin{equation}
O = \sum_{\nu>0} c'_{\nu} a^+_{\nu} a^+_{\bar \nu}.    
\end{equation}
One now rewrites $O$ as 
\begin{equation}
O = \int d^3 r_1 d^3r_2 \chi(r_1,r_2) \psi^+_{\uparrow} (\vec r_1) 
\psi^+_{\downarrow } (\vec r_2) = \sum_{k} \chi(\vec k) a^+_{\vec k \uparrow }
a^+_{\vec k\downarrow},
\end{equation}
where $\chi(\vec k) = V'_k/U'_k$, and define the normalised $N-$particle state  as
\begin{equation}
|N\rangle   =  {\cal N}_N O^{N/2} |0\rangle  
\label{eqA33} 
\end{equation}
and the one-particle density matrix according to 
\begin{equation}
\phi(\vec r_1, \vec r_1 \; ' ) \equiv  
\langle N|\psi^+_{\uparrow}(\vec r_1) \psi_{\uparrow}(\vec r_1 \; ') |N\rangle   =
\langle N| \psi^+_{\uparrow}(\vec r_1) \psi_{\downarrow}(\vec r_1 \;') |N\rangle  .
\end{equation}
Making use of $\psi |0\rangle   = 0$ and of  the commutator  
\begin{equation}
[\psi_{\uparrow}(\vec r_1) , O^{N/2}] = \frac{N}{2} \int d^3 r_2 \chi(\vec r_1,\vec r_2) 
\psi^+_{\uparrow}(\vec r_2) O^{(N-2)/2},
\end{equation}
one can write 
\begin{equation}
\phi(\vec r_1, \vec r_1 \; ') =
\int d^3 r_2 \tilde \chi(\vec r_1  \; ',\vec r_2) \langle N|\psi^+_{\uparrow} (\vec r_1) \psi^+_{\downarrow}(\vec r_2)|N-2\rangle  
\label{eqA28}
\end{equation}
where 
\begin{equation}
\tilde \chi(\vec r_1,\vec r_2) = (N/2){\cal N}_N  {\cal N}^{-1}_{N-2} \chi(\vec r_1,\vec r_2).
\end{equation}
The matrix element in Eq. (\ref{eqA28}) is closely related
with Gorkov's amplitude for two fermions at $\vec r_1$ and $\vec r_2$ to belong to a Cooper pair, i.e.
\begin{equation}
F^>(\vec r_1,\vec r_2) = -i \langle N-2|\psi_{\uparrow}(\vec r_1) \psi_{\downarrow}(\vec r_2)|N\rangle  ,
\end{equation}
its complex conjugate being 
\begin{eqnarray}
F^>(\vec r_1,\vec r_2)^* =
 i \langle N|\psi^+_{\downarrow}(\vec r_2) \psi^+_{\uparrow}(\vec r_1)|N-2\rangle   \nonumber\\
=   -i \langle N|\psi^+_{\uparrow}(\vec r_1) \psi^+_{\downarrow}(\vec r_2)|N-2>.
\end{eqnarray}
Thus, Eq. (\ref {eqA28}) can be written as
\begin{equation}
\phi(\vec r_1, \vec r_1 \; ') = i \int d^3 r_2 \; \tilde \chi(\vec r_1 \; ',\vec r_2) F^\rangle   (\vec r_1,\vec r_2)^*.
 \end{equation}
 Let us now consider the two-particle matrix density 
\begin{eqnarray}
\phi(\vec r_1,\vec r_2; \vec r_3,\vec r_4) \equiv
\langle N| \psi^+_{\uparrow}(\vec r_1) \psi^+_{\downarrow}(\vec r_2)
\psi_{\downarrow}(\vec r_4) \psi_{\uparrow}(\vec r_3)|N\rangle   \nonumber \\
= \phi(\vec r_1,\vec r_3) \phi(\vec r_2,\vec r_4)
 + F^>(\vec r_1,\vec r_2)^* F^> (\vec r_3,\vec r_4),
 \label{eq:twop}
 \end{eqnarray}
equivalent to 
\begin{eqnarray}
 \langle N| \psi_{\uparrow}^+(\vec r_1) \psi^+_{\downarrow}(\vec r_2)
\psi_{\downarrow}(\vec r_4) \psi_{\uparrow}(\vec r_3)|N\rangle    \nonumber\\
= \langle N| \psi_{\uparrow}^+(\vec r_1) \psi_{\uparrow}(\vec r_3)|N\rangle  
\langle N|\psi_{\downarrow}^+(\vec r_2) \psi_{\downarrow}(\vec r_4)|N\rangle   \nonumber \\
+ \langle N| \psi_{\uparrow}^+(\vec r_1) \psi^+_{\downarrow}(\vec r_2)|N-2\rangle  
\langle N-2|\psi_{\downarrow}(\vec r_3) \psi_{\uparrow}(\vec r_4)|N\rangle  .
\end{eqnarray}
The wave function (\ref{eq:twop}) thus leads to a two-particle density matrix fulfilling 
\begin{eqnarray}
{\rm lim}_{\vec r_1,\vec r_2 \to \infty; \vec r_3,\vec r_4 \to -\infty}
\phi(\vec r_1,\vec r_2 ; \vec r_3,\vec r_4) \nonumber \quad \quad (r_{12},r_{34} \langle  \xi),\\
= lim_{\vec r_1 \vec r_2 \to \infty} 
[F^> (\vec r_1, \vec r_2)^*] 
\times
[  {\rm lim} _{\vec r_3 \vec r_4 \to -\infty} F^> (\vec r_3,\vec r_4)] 
\neq 0, 
\end{eqnarray}
property known as ODLRO.

Within the nuclear embodiment, the wave function (\ref{eqA33}) describes 
the properties of a member of a pairing rotation band. For example the ground state of one of the superfluid Sn-isotopes, 
in particular $^{120}$Sn(gs).
In a reaction like $^{120}$Sn + $^{118}$Sn $\to$ $^{118}$Sn(gs) + $^{120}$Sn(gs),
at energies where the distance of closest approach is $2 R_o +a \approx $ 13 fm 
($E_{CM} \approx$ 270 MeV) a number of the effects discussed above can materialise (Fig. A2).
In the tunnelling of a Cooper pair from a superfluid nucleus to the other, each partner 
can be in a different nucleus, but still correlated. This  is in keeping who the fact that the correlation length
arising from the empirical pairing gap ($\Delta \approx 1.4 $ MeV), resulting from the summed
contribution of the bare and induced pairing interaction is
$\xi = \hbar v_F /\pi \Delta  \approx$ 12 fm.

Let us go back to the QRPA (harmonic) diagonalization of $H = H_{MF} + H_p''$. One can
rewrite $H$ as the oscillator \cite{Brink2005},
\begin{equation}
H = \frac{p^2}{2 D''_1} + \frac{1}{2}  D''_1 \omega''_1 q^2
\end{equation}
and identify the momentum with the number operators, the coordinate with the gauge angle and the frequency with the QRPA energy,
\begin{equation}
p = \hbar (\tilde N - N_0), \quad q = \phi, \quad \hbar \omega''_1 = W_1''. 
\end{equation}
The phonon creation operator  for the oscillator is 
\begin{equation}
\Gamma''^+_1 = \sqrt{\frac{\hbar^2}{2D'' W_1''}} (\tilde N - N_0) + i\phi 
\sqrt{\frac{D_1'' W_1''}{2 \hbar^2}}.
\label{eqA38} 
\end{equation}  
Comparing  the coefficient of $(\tilde N - N_o)$ in Eq. (\ref{eqA20}) and (\ref{eqA38}),
and noting that the coefficient of $\phi$ in Eq. (\ref{eqA38}) vanishes in the limit $W''_1 \to 0$, we 
get an expression for the mass parameter 
\begin{equation}
\frac{\hbar^2}{2D''_1W''_1} = \left( \frac{\Lambda''_1}{2 \Delta} \right)^2,
\end{equation}
or 
\begin{equation}
\frac{D''_1}{\hbar^2} = \frac{4 \Delta^2}{2 W''_1 \Lambda''^2_1}.
\end{equation}
Making use of Eq. (\ref{eq:dispers}) one obtains 
\begin{equation}
\frac{1}{W''_1 \Lambda_1''^2}  = 4 \sum_{\nu>0} \frac{2 E_{\nu}}{((2E_{\nu})^2 - (W''_1)^2)^2}.
\end{equation}
In the limit $W''_1 \to 0$ this relation becomes 
\begin{equation}
\frac{1}{W''_1 \Lambda''^2_1} = \sum_{\nu >0}   \frac{1}{2 E^3_{\nu}},
\end{equation}
and the mass parameter  can be written as,
\begin{equation}
\frac{D''_1}{\hbar^2} = \sum_{\nu >0}   \frac{\Delta^2}{ E^3_{\nu}},
\label{eqnoA43}
\end{equation}
emergent property of generalised rigidity  in gauge space for a nucleus whose mean field 
solution violates gauge invariance.

Making use of Eq. (\ref{eqnoA43}), and of the fact that $\lambda = \partial H/\partial N$, the energy of the members 
of a pairing rotational band can be written as 
\begin{equation}
E_N = \lambda N + \frac{\hbar^2}{2 {\cal J}} N^2,
\end{equation}
where 
\begin{equation}
\frac{\cal{J}}{\hbar^2} =\frac{D''_1}{\hbar^2} = \sum_{\nu>0} \frac{4 U_{\nu}^2 V_{\nu}^2} {E_{\nu}} =
2 \sum_{\nu>0} \frac{\langle \nu\bar \nu | \hat N | BCS \rangle  ^2} {2 E_{\nu}},
\end{equation}
is  the cranking formula of the moment of inertia rotations in gauge space. In deriving the above expression use was made of the BCS 
relation $2 U_{\nu} V_{\nu} = \Delta/E_{\nu}$.
Pairing rotations can be viewed as the Goldstone--mode, or better the Anderson--Goldstone--Nambu mode
\cite{Bes1966,Hinohara2016,Broglia2000,Lopez:13} in gauge space,  approaching the $E=0$ limit linear with $N$. This is in keeping with the fact that such
behavior  is only expected in the laboratory system, where it can be measured. In other words, summing to the BCS energy 
$U$ the Coriolis force $\lambda N$ in gauge space  felt by the condensate in the intrinsic system.

\begin{figure}[h!]
\centerline{\includegraphics[width=0.2\textwidth]{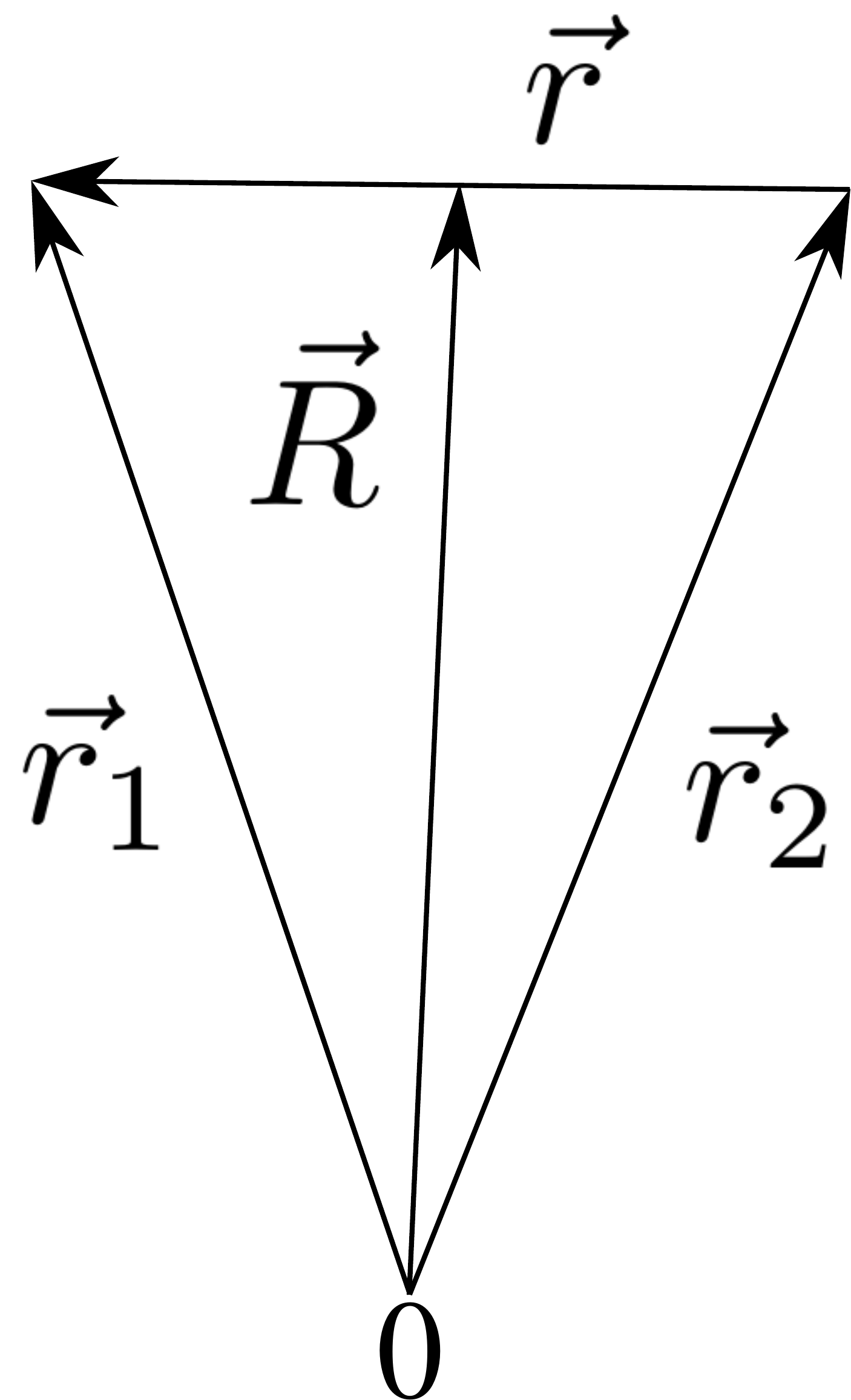}}
\caption{Coordinates used to define the pair operator $P^+(\vec R)$.}
\end{figure}

\begin{figure}[h!]
\centerline{\includegraphics[width=0.5\textwidth]{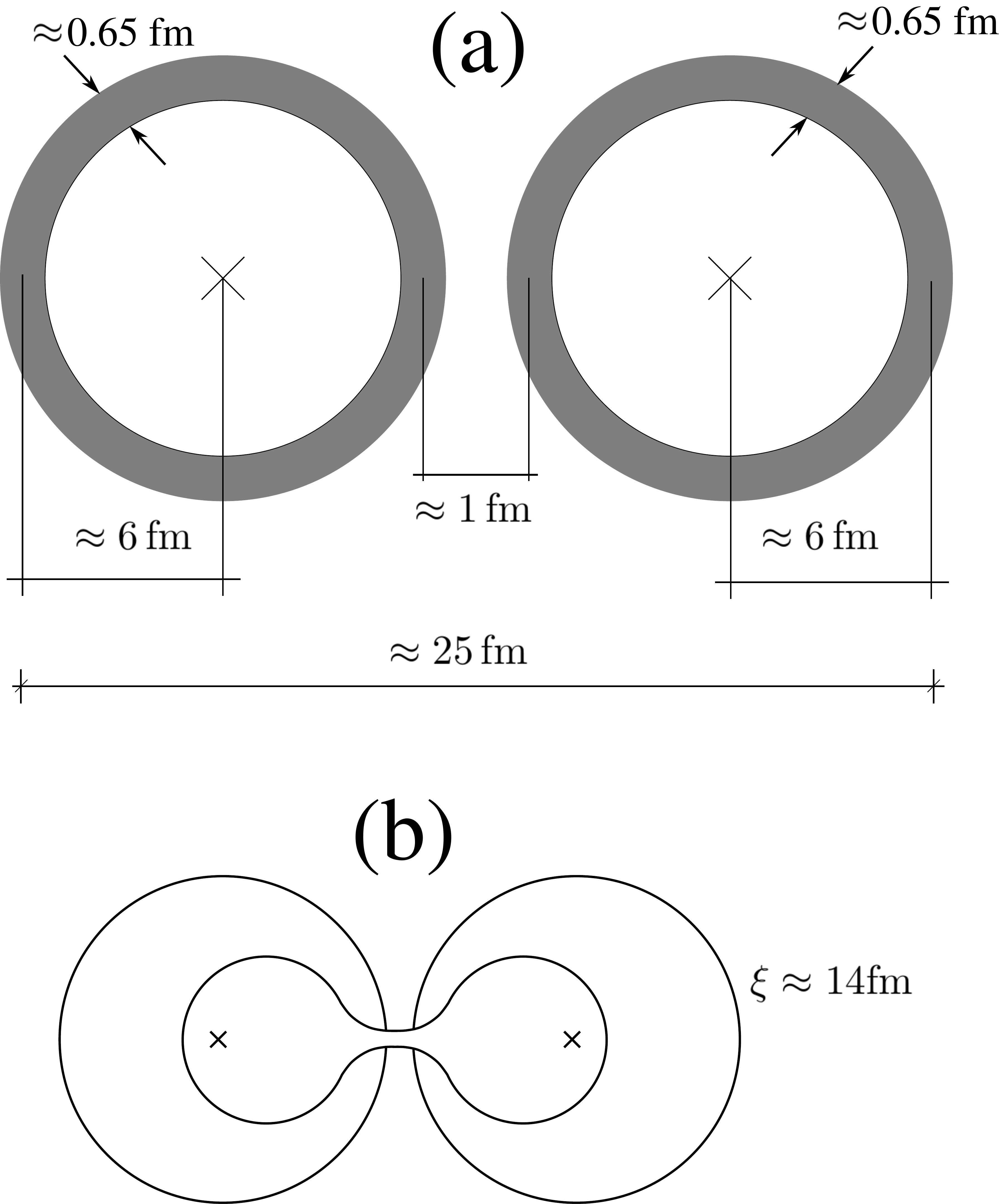}}
\caption{(a) Schematic representation of two Sn nuclei 
at a distance of closest approach of $\approx 13$ fm. (b) Single 
Cooper pair in which each nucleon is in a different nucleus.}
\end{figure}

\clearpage
\setcounter{equation}{0}
\appendix 
\chapter{\large \bf  Appendix B. Useful definitions}

\vspace{1cm} 

\makeatletter
\renewcommand{\theequation}{B\@arabic\c@equation}
\makeatother

\makeatletter
\renewcommand{\thefigure}{N\@arabic\c@figure}
\makeatother

Making use of the  wave function $u_{\nu}(\vec r, \sigma)$,
position and  spin representation of the ket $|\nu\rangle  $ describing the single-particle  motion of a nucleon in the state  
$\nu$,
\begin{equation}
u_{\nu} (\vec r , \sigma) = \langle  \vec r, \sigma | \nu\rangle  
\end{equation}
one can define the field operator
\begin{equation}
\psi(\vec r, \sigma) = \sum_{\nu} \langle \vec r, \sigma| \nu\rangle  a_{\nu} =
\sum_{\nu} u_{\nu} (\vec r, \sigma) a_{\nu},
\label{eqB2}
\end{equation}
$a_{\nu}$ being the  annihilation operator  of a fermion in state  $\nu$.
Thus, $\psi(\vec r ,\sigma)$ is an operator in the occupation-number space and also a function of position and spin, while 
 \begin{equation}      
\psi^+ (\vec r, \sigma) = \sum_{\nu} u_{\nu}^* (\vec r ,\sigma) a^+_{\nu}
\label{eqB3}
\end{equation}
is its hermitian conjugate, $a^+_{\nu}$ being the creation operator  of a fermion in state $\nu$, i.e.
$|\nu\rangle   = a^+_{\nu}|0\rangle  $, $|0\rangle  $ being the vacuum state.

A one-particle operator, e.g. the one-particle potential can be written as,
\begin{equation}
U = \sum_{\nu,\nu'} \langle \nu|V|\nu'\rangle   a^+_{\nu} a_{\nu'},
\end{equation}
where 
\begin{eqnarray}
\langle \nu |U |\nu'\rangle   = \sum_{\sigma,\sigma'} \int d^3 r \langle  \nu | \vec r, \sigma\rangle    \langle \vec r, \sigma |U | \vec r, \sigma\rangle  
\langle \vec r ,\sigma'| \nu'\rangle   \nonumber  \\
= \sum_{\sigma,\sigma'} \int d^3 r \;  u^*_{\nu} (\vec r, \sigma) V_{\sigma,\sigma'} (\vec r) u_{\nu'} (\vec r, \sigma').
\end{eqnarray}
Thus,
\begin{equation}
U = \sum_{\sigma,\sigma'} \int d^3 r \sum_{\nu,\nu'} u^*_{\nu} (\vec r,\sigma) U_{\sigma,\sigma'}(\vec r) u_{\nu'}(\vec r, \sigma')
a^+_{\nu} a_{\nu'}.
 \end{equation}
Making use  of Eq. (\ref{eqB2}) and (\ref{eqB3}) one can write, 
\begin{equation}
U = \sum_{\sigma,\sigma'} \int d^3 \psi^+ (\vec r, \sigma) U_{\sigma,\sigma} \psi(\vec r ,\sigma').
\end{equation}
Let us consider  now the two-body interaction 
\begin{equation}
v = - \frac{1}{2} \sum_{\nu_1 \nu_2 \nu_1 ' \nu_2 ' } v_{\nu_1 \nu_2, \nu_1 ' \nu_2 '} 
a^+_{\nu_2} a^+_{\nu_1} a_{\nu_1 '} a_{\nu'_2} ,
\end{equation}
where 
\begin{eqnarray}
v_{\nu_2\nu_1, \nu_1'\nu_2'} = \langle \nu_2 \nu_1 |v| \nu_1'\nu_2'\rangle  
= \int d^3 r_1 d^3 r_2 \sum_{\sigma_1,\sigma_2,\sigma_1'\sigma_2'} \langle \nu_2 | \vec r_2, \sigma_2 \rangle  \langle \nu_1 | \vec r_1,\sigma_1\rangle   \nonumber  \\
\times \langle \vec r_2,\sigma_2; \vec r_1,\sigma_1 |v| \vec r_1,\sigma'_1; \vec r_2, \sigma_2 \rangle   \langle \vec r_1,\sigma'_1|\nu'_1\rangle   \langle \vec r_2,\sigma'_2| \nu'_2\rangle   \nonumber\\
= \sum_{\sigma_1,\sigma_2,\sigma'_1,\sigma'_2} \int d^3 r_1 d^3 r_2 u^*_{\nu_2}(\vec r_2,\sigma_2 ) u^*_{\nu_1} (\vec r_1,\sigma_1) 
v_{\sigma_2 \sigma_1,\sigma'_1\sigma'_2} (\vec r_1,\vec r_2)  \nonumber \\
\times u_{\nu'_1} (\vec r_1,\sigma'_1) u_{\nu'_2} (\vec {r}_2'.\sigma'_2).
\end{eqnarray}
Thus 
\begin{eqnarray}
v =  - \frac{1}{2}  
\sum_{\sigma_1,\sigma_2,\sigma'_1,\sigma'_2} \int d^3 r_1 d^3 r_2 \nonumber  \\
\sum_{\nu_1,\nu_2,\nu'_1,\nu'_2} 
u^*_{\nu_2}(\vec r_2,\sigma_2) u^*_{\nu_1} (\vec r_1,\sigma_1)  
v_{\sigma_2 \sigma_1,\sigma'_1\sigma'_2} (\vec r_1,\vec r_2) \nonumber  \\
\times u_{\nu'_1} (\vec r_1 \;',\sigma'_1) 
u_{\nu'_2} (\vec  r_2\;',\sigma'_2)  a^+_{\nu_2} a^+_{\nu_1} a'_{\nu_1} a_{\nu_2}'  \nonumber  \\
=  \frac{1}{2}  
\sum_{\sigma_1,\sigma_2,\sigma_1 ' ,\sigma_2 '} \int d^3 r_1 d^3 r_2 
\psi^+(\vec r_2,\sigma_2) \psi^+(\vec r_1,\sigma_1)  \nonumber  \\
\times v_{\sigma_2 \sigma_1,\sigma_1 ' \sigma_2 '} (\vec r_1,\vec r_2) 
\psi(\vec r_2,\sigma_2 ') \psi(\vec r_1,\sigma_1 ').
\end{eqnarray}

\clearpage
\setcounter{equation}{0}
\appendix 
\chapter{\large \bf  Appendix C. One- and two-body Dirac matrices}

\vspace{1cm} 

\makeatletter
\renewcommand{\theequation}{C\@arabic\c@equation}
\makeatother

\makeatletter
\renewcommand{\thefigure}{N\@arabic\c@figure}
\makeatother

Short before the appearance of the BCS papers \cite{Bardeenetal1957a,Bardeenetal1957b}, the result
of a study concerning the nature of the order parameter of a boson superfluid, such as $^4$He below 2.17 K, was published 
\cite{Penrose}. In this reference it was argued that the Bose condensation that is supposed to be responsible for superfluidity 
should manifest itself in the off-diagonal elements of the one-particle density matrix 
\begin{equation}
\rho(\vec r, \vec r\;') = \langle \phi^+(\vec r\;') \phi(\vec r )\rangle  ,
\label{eqc1}
\end{equation}
where $\phi^+ (\phi) $ are boson creation (annihilation) operators, the expectation 
value being taken in some statistical ensemble of states. This Dirac density matrix is Hermitian, 
its trace giving the total number of bosons $N$. In the normal state its eigenvalues are all at most
of order unity, but in the superfluid state there is a macroscopic eigenvalue $n_o$, much larger 
than one. 
In the case of superfluid  helium ($^4$He) $n_o/N$ seems to be  of the order of 10\%, $N$ being the total number of atoms. In other
words when $\vec r $ and $\vec r\;'$ are far apart $\rho(\vec r ,\vec r\;')$ must tend to $n_o/N$ in the superfluid 
state. This is a feature which is very different from that of a solid e.g. argon at low temperature, and is due to the high zero point 
motion of the light helium atoms, and the weak interaction between them. In the case of solid argon, the removal of one 
particle from its equilibrium position and transport to a distant point creates a vacancy-interstitial pair, the energy barrier associated 
with such process makes the amplitude of off-diagonal elements of the one-particle 
density matrix to fall off exponentially with the separation of the two coordinates.

In the case of fermions such as electrons or $^3$He atoms, the one-particle density matrix cannot have a macroscopic eigenvalue, since its 
eigenvalues lie in the range [0,1], but the two-particle density can \cite{Yang1962}. The two-particle Dirac density matrix 
has the form,
\begin{equation}
\Gamma(\vec R', \vec r\;'; \vec R, \vec r) = \langle \psi^+(\vec R\;' + \frac{\vec r\;'}{2}) \psi^+(\vec R' - \frac{\vec r\;'}{2})
\psi(\vec R -  \frac{\vec r}{2}) \psi(\vec R + \frac{\vec r}{2}),
\end{equation}
where,  for simplicity, the spin variables for the fermion creation and annihilation operators $\psi^+,\psi$ have been suppressed. 
Disregarding, again for simplicity, crystal structure and disorder of the lattice, electrons move in a translational 
invariant environment. In that case ,the eigenvectors of the matrix depend on the center of mass coordinates 
only through factors of type $exp(i \vec K \cdot \vec R$). In the BCS ground state,
\begin{equation}
\Gamma= \delta(K',0) \delta(K,0) U_{k'}V_{k'}U_k V_k.
\end{equation}
In configuration space the eigenvector corresponding to the macroscopic eigenvalue 
is the Fourier transform of this factor, so that the order parameter of the ground state is constant in $\vec R$ and $\vec R\;'$, and is spread 
out in its internal coordinates $\vec r \;$ and $\vec r\; '\, $ by an amount which depends  on how strongly peaked the coherence 
fact is about the Fermi surface. This dependence on $\vec r$ gives the wave function of the Cooper pair.

In other words, and in keeping with the fact that, according to Wick's theorem,  
\begin{eqnarray}
\Gamma(\vec r_1,\vec r_2; \vec r_1\;',\vec r_2\;')  = \langle \psi^+(\vec r_2\;') \psi^+(\vec r_1\;') \psi(\vec r_1) \psi(\vec r_2) \rangle   = \nonumber \\
\rho(\vec r_2,\vec r_1\;') \rho(\vec r_2, \vec r_2\;') - \rho(\vec r_1,\vec r_2\;') \rho(\vec r_2,\vec r_1\;') +
\chi^*(\vec r_2\;',\vec r_1\;') \chi(\vec r_2,\vec r_1),
\end{eqnarray}
where $\chi^*(\vec r_1,\vec r_2) = \langle \psi(\vec r_1)\psi(\vec r_2)\rangle  , $ the eigenvector of $\Gamma$ being the (pairing) function $\chi$. That is ,
\begin{equation}
\int \int  \Gamma(\vec r_1,\vec r_2; \vec r_1', \vec r_2) \chi(\vec r_1\;' \vec r_2\;') d^3 r_1' d^3 r_2' \approx n_0 \chi(\vec r_1,\vec r_2),
\end{equation}
where the eigenvalue, namely the pair density, is large, and is related to the large overlap of Cooper pairs at the basis of BCS theory. 
Within this context, the coherence length of a metallic superconductor being of the order of $10^4 \AA$, implies that there are of the order 
of $10^{11}$ the electrons within a coherence volume, in keeping with the fact an electron typically occupies a volume $\approx (2 \AA )^3$
(Wigner-Seitz cell).

In a neutron superfluid nucleus like e.g. $^{120}$Sn, close to   15\% of all nucleons participate in the  condensate, the number of Cooper pairs being of the 
order of 5-6.

\vspace{1cm} 

\makeatletter
\renewcommand{\theequation}{D\@arabic\c@equation}
\makeatother

\makeatletter
\renewcommand{\thefigure}{D\@arabic\c@figure}
\makeatother

\newpage
\setcounter{equation}{0}
\appendix 
\chapter{\large \bf  Appendix D. NFT vacuum polarization}

\vspace{1cm} 

\makeatletter
\renewcommand{\theequation}{D\@arabic\c@equation}
\makeatother

\makeatletter
\renewcommand{\thefigure}{D\@arabic\c@figure}
\makeatother

The role zero point fluctuations play in the nuclear ground state, i.e. in the NFT vacuum  can
be  clarified by relating it to the polarisation of the QED vacuum.
Let us briefly dwell on the "reality" of such phenomenon by recalling the fact that
to the question of Rabi of whether the polarisation of the QED vacuum could be measured  \cite{Pais} - in particular
the change in charge  density felt by the electrons of an atom,  e.g. the electron of a hydrogen atom, due to
virtual creation and annihilation of electron-positron pairs - Lamb gave a quantitative answer, both experimentally
and theoretically \cite{Lamb,Kroll}. The corresponding correction  (Lamb shift) implies that the $2s_{1/2}$ level lies 
higher than the $2p_{1/2}$ level by about 1000 megacyles/s  as experimentally observed.

In connection with the discussion of Feynman of vacuum polarisation, where a field produces a pair,
the subsequent pair annihilation producing a new field, namely a close loop, he implemented in his space--time trajectories 
Wheeler's idea of electrons going backwards in time (positrons).  Such trajectories would be like an $\mathbf N$ in time,
that is electrons which would back up for a while, and go forward again. Being connected 
with a minus sign, these processes are associated with Pauli principle in the self--energy of electrons
(see Fig. 1,I(c)). 
The divergences affecting
such calculations  could be renormalised by first computing the self-energy  diagram in second order and finding the answer which is finite, but contains a cut-off
to avoid a logarithmic divergence. Expressing the result in terms of the experimental mass, one can take 
the limit (cut-off $\to \infty$) which now exists.
Concerning radiative corrections to scattering, in particular that associated with the process
in which the potential creates an electron-positron pair which then reannihilates, emitting a quantum which scatters the
electron, the renormalisation procedure should be applied to the electric charge, introducing 
the observed one (Bethe and Pauli, see \cite{Bethe}).

In the nuclear case, for example Skyrme effective interactions give 
rise to particle-vibration coupling vertices which, because of the contact character of these interactions 
may lead to divergent zero point energies, unless a cut-off is introduced\footnote{Let alone the fact that the velocity dependent component of these forces weaken the PVC vertices leading to poorly collective low--lying vibrations, and to equally poor clothed valence states. The question emerges of which are the provisos to be taken in the use of effective forces to higher orders of the PVC. Within this context cf. \cite{Mahaux}, also \cite{Physica_scripta,11Be} concerning the implementation of renormalization in both configuration and 3D--spaces within the framework of NFT. In a nutshell, the bare mean field exists but its properties cannot be measured (not any more than the bare electron mass in renormalized quantum electrodynamics), and corresponds to a set of parameters of a Fermi--like function which ensure that the clothed states reproduce all of the experimental findings, both structure and reaction.}.
The Gogny force being finite range does not display such problems. Nonetheless, 
the associated results concerning zero point energies may not be very stable and/or accurate 
carrying out a complete summation over both collective and non collective contributions. 
In this case one can eliminate such a  problem by going to higher orders in the oyster diagrams (see Fig. 1(I)(a)). 
The fermion exchange between two of these diagrams (Pauli principle) essentially eliminates all of the non-collective
contributions, leading to accurate results.

An economic and quite reliable method to achieve a similar result, 
is that of using empirical renormalisation. That is, to calculate the lowest order diagrams 
but introducing, in the intermediate states, the dressed physical (empirical) states 
(\cite {11Be,Mattuck}; see also \cite{Bes1977}).

\newpage
\setcounter{equation}{0}
\appendix 
\chapter{\large \bf  Appendix E. State dependent effective mass and mean field potential}

\vspace{1cm} 

\makeatletter
\renewcommand{\theequation}{E\@arabic\c@equation}
\makeatother

\makeatletter
\renewcommand{\thefigure}{E\@arabic\c@figure}
\makeatother

The bare  mass of a nucleon in the nucleus is not a quantity  that can be measured. This is
because a nucleon in the nucleus is subject to a mean field which is both non-local in space as
well as in time.

The first  component arises already at the level of Hartree-Fock, and is directly related to the Hartree exchange potential,
assuming velocity independent interactions. This non locality can be taken care of,  in most situations, 
in terms of an effective mass, the $k$-mass, its average value being $m_k \approx 0.7 m $, where  $m$
is the bare mass. The quantity $m_k$ is intimately related to the so called Perey-Buck potential, namely the energy dependent 
term in the strength $V = V_0 + 0.4 E$ of the real part of the optical potential 
needed to describe nucleon-nucleus elastic scattering  experiments at bombarding energies of tens of MeV, where
$E= |\epsilon_k - \epsilon_F| (\epsilon_k = \hbar^ 2 k^2 / 2 m)$. One can obtain essentially the same results by solving 
the elastic scattering single-particle Schr\"odinger equation making use of an energy independent potential of strength 
$V \approx \left( m/m_k \right) V_0 = 1.4 V_0$ and of an effective mass $m_k= \left( m (1 + (m/(\hbar^2 k) dV/dk \right)^{-1}$
(within this context see Fig. 2.14 in \cite{Mahaux}). Similar results and protocol are obtained and can be used 
to describe deep hole states. 

In other words, the concept of a single, mean field potential is a somewhat illusory one. This is even more so in keeping
with the fact that there is  not a single $k-$mass, but a state dependent one equal to the expectation value of the 
quantity in parenthesis, where $V$ is now the sum of the direct and exchange potential, calculated making use of the corresponding 
single-particle wave functions \cite{Bernard}.

Retardation effects arise from the coupling of single-particles with collective vibrations (Fig. 1(I)). They lead,
for states close to the Fermi energy, to the state-dependent $\omega-$mass ($m_{\omega} = m (1+ \lambda),
Z_{\omega} = m/m_{\omega}$),  and to fragmentation, effects which can hardly be parameterised in terms of an average 
mean field potential. 

In other words, the above effects are at the basis of the dynamical shell model. While one can, within this context, 
accurately
calculate the single-particle properties ($\tilde \epsilon_{a(n)},Z_{a(n)},N_{a(n)})$ in simple and economic ways,e.g.  renormalized
NFT, the situation is much more complex  concerning the absolute value of the Fermi energy. 
\setcounter{equation}{0}
\appendix 
\chapter{\large \bf  Appendix F. Correlation length, correlation energy,
generalised quantity parameter }

\vspace{1cm} 

\makeatletter
\renewcommand{\theequation}{F\@arabic\c@equation}
\makeatother

\makeatletter
\renewcommand{\thefigure}{F\@arabic\c@figure}
\makeatother

The correlation length is defined as (\cite{Schrieffer1964}, p.18)
\begin{equation}
\xi =\frac{\hbar v_F}{\pi \Delta},
\end{equation}
while the condensation energy  i.e. the difference between the ground state energy of the 
normal $(N)$ and superfluid $(S)$ phases is (\cite{Schrieffer1964}, eq. (2-35))
\begin{equation}
E_{cond} =W_N - W_S = \frac{1}{2} N(0) \Delta^2.
\label{cond}
\end{equation}
In the above equations $v_F$ is the Fermi velocity , $\Delta$ the pairing gap,
and $N(0)$ the density of levels at the Fermi energy for one spin orientation. 
The correlation energy $E_{corr}$ introduced in \cite{Bohr1975} (Eq. (6-618)) has the
opposite sign to (\ref{cond}). Concerning the density of levels of the corresponding  reference,
a spectrum of equally spaced levels, each of them displaying  two-fold Kramers' degeneracy was used,
typical of quadrupole deformed nuclei (Nilsson model). Calling $d$ the spacing between them
($d \approx$ 0.4-0.5 MeV), the density of levels for both spin orientation is $2/d$, while
$N(0)= 1/d$. Thus, $E_{corr} = - \Delta^2/2d$ coincides in absolute value with  (\ref{cond}),
as expected.

Making use of the empirical values for the density of levels of both spin orientations,
namely $a= N/8$ MeV$^{-1}$ (\cite{Bortgdr}, eq. (7.16)) one obtains 
for $^{120}_{50}$Sn$_{70}$,
$N(0) \approx$ 4 MeV$^{-1}$. It is of notice that for this nucleus 
the empirical  value of     the pairing gap is $\Delta \approx 1.45 $ MeV. The associated
correlation length amounts to $\xi \approx 12$ fm. That associated 
with $\Delta_{ind} = (1/2) \Delta$  being of course 24 fm, value appearing in Eq. (\ref{xivalue}).
Thus $E_{corr} \approx -$ 5 MeV, while the value associated
with
$\Delta_{ind} (\approx 0.8 $ MeV) is -1.3 MeV.   
Making use of the above values, the generalised quantality parameter is, 
\begin{equation}
q_{\xi}=  \frac{\hbar^2}{2m \xi^2} \frac{1}{|E_{corr}|} =  0.03 \; (\xi = 12 \; {\rm fm},  E_{corr} = -5 \; {\rm MeV} ),
\end{equation}
and testifies to a strong correlation between the partners of the Cooper pair.

\setcounter{equation}{0}
\appendix 
\chapter{\large \bf  Appendix G. Induced pairing interaction}

\vspace{1cm} 

\makeatletter
\renewcommand{\theequation}{G\@arabic\c@equation}
\makeatother

\makeatletter
\renewcommand{\thefigure}{G\@arabic\c@figure}
\makeatother

\begin{figure}[h!]
\centerline{\includegraphics[width=0.6\textwidth]{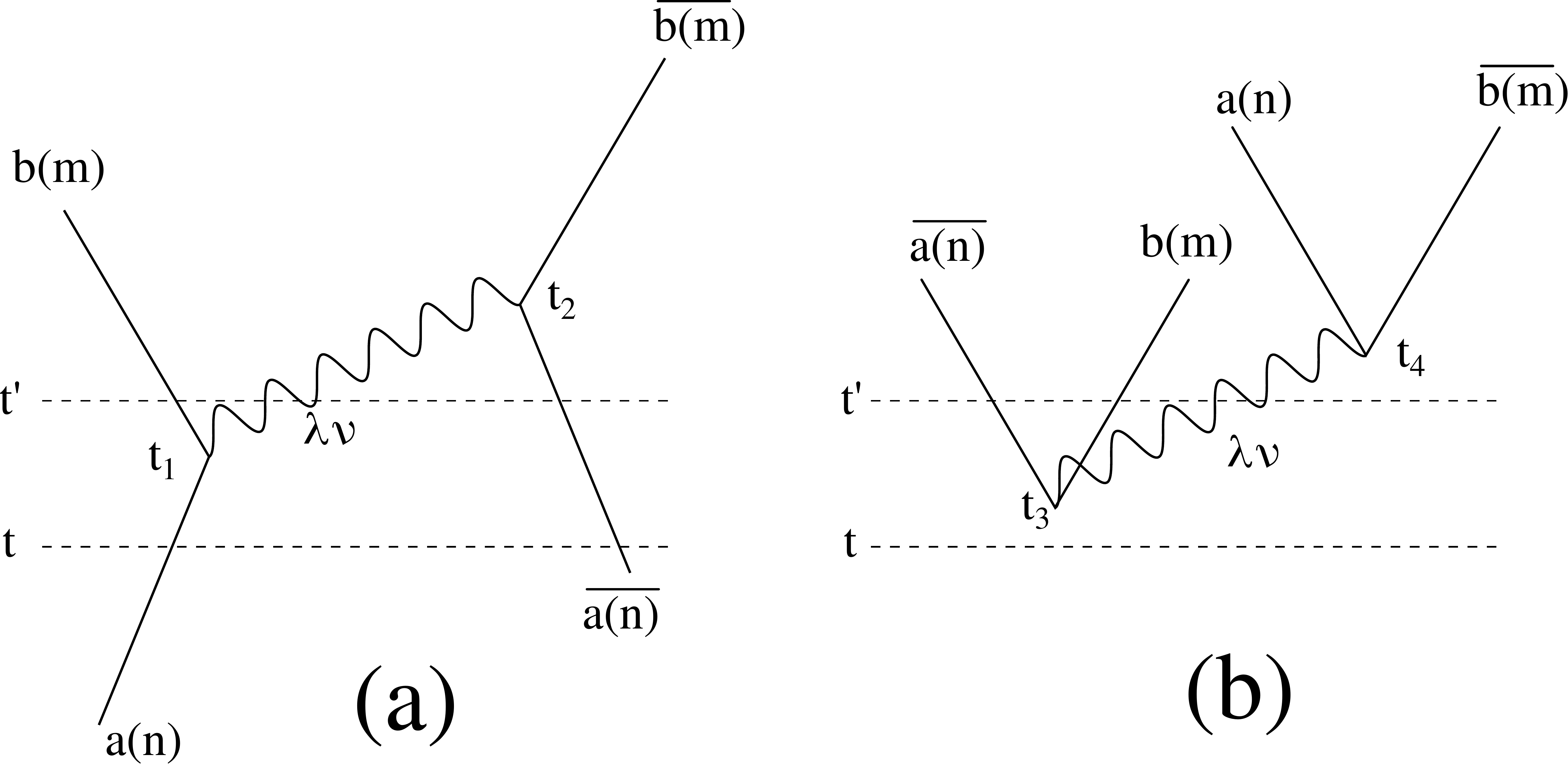}}
\caption{Schematic representation of the induced pairing interaction (\ref{eq28}).}
\label{figG1}
\end{figure}

The exchange of collective vibrations between nucleons moving in time reversal states give rise to an induced, medium polarization, pairing interaction. In the quasiparticle representation and QRPA treatment of the collective modes, the different lowest order contributions in the PVC vertices to (\ref{eq28}) are shown in Fig. \ref{figG1} (see Fig \ref{fig2} (e)--(g) for examples of higher order).
To each vertex is associated a function $h(a(n),b(m)\lambda\nu)$. The denominator corresponds to the energy difference between the configuration at time $t$ and at time $t'$, i.e.
\begin{equation}
\textbf{(a)}=\sum_{\lambda\nu}\frac{2|h(a(n),b(m)\lambda \nu)|^2}{\tilde E_{a(n)}-\tilde E_{b(m)}-\hbar\omega_{\lambda\nu}} \quad\textrm{and}\quad \textbf{(b)}=-\sum_{\lambda\nu}\frac{2|h(a(n),b(m)\lambda \nu)|^2}{\tilde E_{a(n)}+\tilde E_{b(m)}+\hbar\omega_{\lambda\nu}},
\end{equation}
the factor of 2 arising from the two time ordered contributions, i.e. $t_1<t_2$ and $t_2>t_1$, and $t_3<t_4$ and $t_4>t_3$ respectively. Non--arrowed lines represent quasiparticles, the wavy line QRPA vibrations of multipolarity $\lambda$ and increasing energy labeled by $\nu$. They contain both particle and hole components, and one has to consider both contributions simultaneously. 
\setcounter{equation}{0}
\appendix 
\chapter{\large \bf  Appendix H. Vertex corrections}\label{AppG}

\vspace{1cm} 

\makeatletter
\renewcommand{\theequation}{H\@arabic\c@equation}
\makeatother

\makeatletter
\renewcommand{\thefigure}{H\@arabic\c@figure}
\makeatother

The PVC mechanism gives rise to self energy processes (e.g. Fig. 1(II)(a),(c)(d)) but also
to vertex renormalisation (Fig. 1(II)(e)). In other words, $h(a,b\lambda\nu)$ (Fig. H1 (a)), is to be
corrected to lowest order in the PVC vertex (Fig.  H1(b)), correction which can be written as
\begin{equation}
\delta h (a,b\lambda\nu) = \sum_{c,\lambda'\nu'}  \frac{Q(b\lambda\nu;c\lambda'\nu') h(a,c\lambda'\nu')} {\tilde E_a - (E_c +\hbar \omega_{\lambda'\nu'})}
\label{eq:vert}
\end{equation}
where (Fig H1(c)),
\begin{equation}
Q(b\lambda\nu;c \lambda'\nu') = \sum_{d} h(b, d\lambda' \nu') 
\frac{\langle (j_d\lambda')j_b;j_a|(j_d\lambda)j_c,\lambda';j_{a}\rangle  }{\tilde E_a - (E_{d} + \hbar \omega_{\lambda}+ \hbar \omega_{\lambda'\nu'})}
h(c,d\lambda\nu),
\label{eq:recoup}
\end{equation}
$\langle (j_d\lambda')j_b;j_a|(j_d\lambda)j_c,\lambda';j_{a}\rangle  $ being a recoupling coefficient.

Correction (\ref{eq:vert}) with the proper indexing, has to be added to both vertices entering the expression for $v_{ind}$,
as well as to those of the various self energies.
In the case under discussion namely $^{120}$Sn, that is a medium heavy superfluid nucleus lying along the stability valley,
the recoupling coefficient (Eq. (\ref{eq:recoup})) displays rather random phases leading to strong  cancellations
when summed over the different quantum numbers, resulting in values of $\delta h$ of the order of few tens of keV.

Similar arguments apply to the bare pairing interaction vertex correction (Fig. \ref{fig:vert_beta}), and to the Dyson equation \cite{Idini_tesi}.

\begin{figure}[h!]
\centerline{\includegraphics[width=0.6\textwidth]{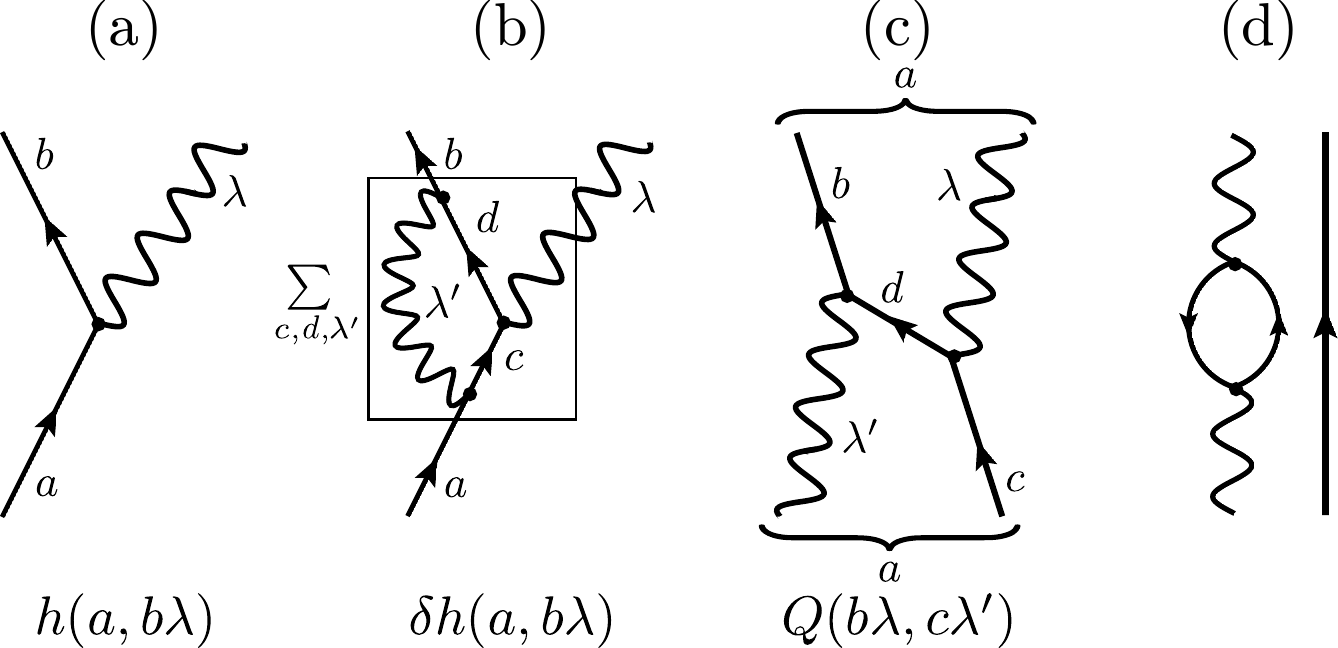}}
\caption{Particle-vibration coupling  (PVC) vertex renormalisation of the induced pairing interaction. 
{(\bf a)}  PVC $h(a,b\lambda\nu)$. {(\bf b)} renormalization; {(\bf (c)} the process boxed 
in diagram (b) corresponding to one of those describing the Compton effect in quantum electrodynamics, 
and resulting from the time ordering of the Pauli principle correction between the single nucleon 
considered explicitly and those out which the vibrations are built, and shown in {(\bf d)}.}
\label{fig:vert_alpha}
\end{figure}

\begin{figure}[h!]
\centerline{\includegraphics[width=0.5\textwidth]{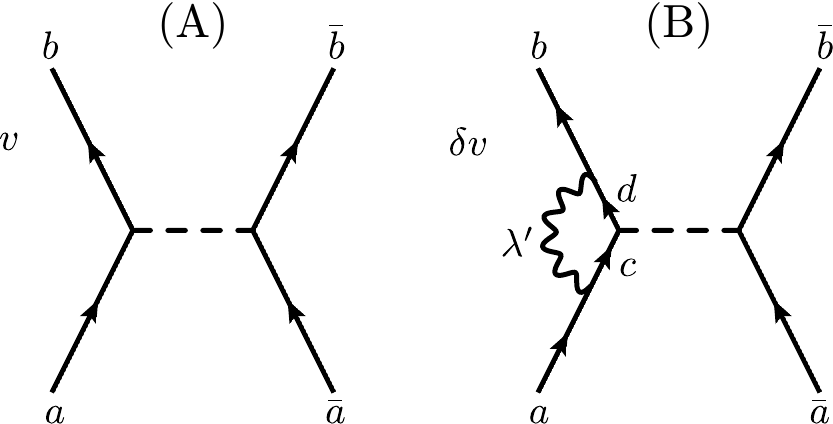}}
\caption{Particle-vibration coupling  (PVC) vertex renormalisation of the bare pairing interaction. 
{(\bf A)}  Process through which two nucleons moving in time reversal states 
interact through the bare pairing interaction.  {(\bf (B)} Vertex renormalisation 
through the PVC.}
\label{fig:vert_beta}
\end{figure}


\clearpage


\newpage 

\begin{thebibliography}{}
\bibitem{Mottelson1998}  B.R. Mottelson, in {\it Trends  in nuclear physics. 100 years later}, Les Houches,
Session LXVI, Elsevier (Amsterdam) p. 25 (1998)
\bibitem{Brink2005} D.M. Brink and  R.A. Broglia,  
{\it  Nuclear superfluidity}, Cambridge University Press, Cambridge  (2005) 
\bibitem{Bohr1975}  A Bohr and B.R. Mottelson,   {\it  Nuclear structure },  Vol II, Benjamin (New York) (1975)
\bibitem{Bes1966}  D.R. Bes and R.A. Broglia,  {\it Pairing vibrations}, Nucl. Phys.  {\bf 80} (1966) 289
\bibitem{Broglia2016} R.A. Broglia, {\it Nuclear structure then and now: 40 years of the 1975 Nobel prize in physics}, Ann. Phys.  {\bf 528} (2016) 444
\bibitem{Heisenberg1932} W. Heisenberg, {\it About the constitution of atomic nuclei. 1}, Zeit Phys. {\bf 77} (1932) 1
\bibitem{Mayer1950} M.G. Mayer, {\it Nuclear configurations in the spin-orbit coupling model. 2.Theoretical considerations},
Phys. Rev. C {\bf 78}  22 (1950) 
\bibitem{Racah1952}  G. Racah and I.Talmi, {\it The pairing properties of nuclear interactions}, Physica {\bf 18}, 1097 (1952)
\bibitem{Bardeenetal1957a}  J. Bardeen, L. N. Cooper and J.R. Schrieffer, {\it Microscopic theory of superconductivity}.  Phys. Rev. {\bf 106 }162 (1957)
\bibitem{Bardeenetal1957b}  J. Bardeen, L. N. Cooper and J.R. Schrieffer, {\it Theory of superconductivity}.  Phys. Rev. {\bf 108 }1175 (1957)
\bibitem{Bohr1958}  A. Bohr, B.R. Mottelson and D. Pines, {\it Possible analogy between the excitation spectra of nuclei and those of the superconducting metallic state},
Phys. Rev .{\bf 110} (1958) 936 
\bibitem{Bohr1964} A. Bohr,   {\it  Elementary modes of excitation and their coupling },
Comptes Rendues du Congres International de physique nucl�aire, CNRS, 487 (1964)
 \bibitem{Yoshida1962} S. Yoshida,  {\it  Note on the two-nucleon stripping reaction}
 Nucl. Pays.  {\bf 33}, 685 (1962)
\bibitem{Poteletal2013} G. Potel, A. Idini, F. Barranco, E. Vigezzi and R.A. Broglia,  {\it Quantitative study of coherent pairing modes with two-neutron transfer: Sn isotopes}, Phys. Rev. C {\bf 87} (2013) 054321
\bibitem{Poteletal2013a} G. Potel, A. Idini, F. Barranco, E. Vigezzi and R.A. Broglia,  {\it Cooper pair transfer in nuclei}, 
Rep. Prog. Phys. {\bf 76} (2013) 106301
\bibitem{EPJ}  F. Barranco, R.A. Broglia, G. Col\`o, G. Gori,  E. Vigezzi and P.F. Bortignon, {\it Many-body effects in nuclear structure}, Eur. J. Phys. A {\bf 21}, 57 (2004) 
\bibitem{Idini2015}  A. Idini, G. Potel, F. Barranco, E. Vigezzi and R.A. Broglia, {\it Interweaving of elementary modes of excitation 
in superfluid nuclei through particle-vibration coupling: quantitative account of the variety of nuclear excitations}, Phys. Rev C {\bf 92}  (2015) 014331 
\bibitem{Bortgdr} A. Bracco, P.F. Bortignon and R.A. Broglia,  {\it Giant resonances}, Harwood Academic Publishers , Amsterdam (1998) 
\bibitem{Mahaux} C.Mahaux, P. F. Bortignon, R.A.Broglia, C. H. Dasso,
{\it Dynamics of the Shell Model}, Phys. Reports {\bf 120} (1985) 1
\bibitem{Baroni} S. Baroni, M. Armati, F. Barranco, R.A. Broglia, G. Col\`{o}, G. Gori  and  E. Vigezzi,
{\it Correlation energy contribution to nuclear masses}, J. Phys. G {\bf 30} (2004) 1353  
\bibitem{Born} M. Born, {\it Natural philosophy of cause and chance}, Clarendon Press, Oxford (1949)
\bibitem{Schwinger} J. Schwinger, {\it Quantum electrodynamics}, Dover, New York (1958)
\bibitem{Feynman}  R.P. Feynman, {\it Quantum electrodynamics}, Benjamin, New York (1962)
\bibitem{Broglia2000} R.A. Broglia,  J. Terasaki and  N. Giovanardi ,
{\it  The Anderson-Goldstone-Nambu mode in finite and infinite systems}, Phys. Rep.  {\bf 335} (2000) 1
\bibitem{Hinohara2016} N. Inohara and W. Nazarewicz,  
{\it  Nambu-Goldstone modes with nuclear density functional theory}, Phys. Rev. Lett.
 {\bf 116} (2016) 152502
 \bibitem{BrogliaHansenRiedel} R.A. Broglia, O. Hansen and C. Riedel,  {\it  Two-neutron
transfer reactions and the pairing model }, Adv. Nucl. Phys. {\bf 6}, 287 (1973) (see \url{www.mi.infn.it/~vigezzi/BHR/ BrogliaHansenRiedel.pdf}) 
\bibitem{Gorkov} L.P. Gor'kov,  {\it  On the energy spectrum of superconductors} Sov. Phys. JETP {\bf 7} (1958) 505
\bibitem{Nambu}  Y. Nambu, \textit{Quasi--Particles and Gauge Invariance in the Theory of Superconductivity} Phys. Rev. {\bf 117} (1960) 648
\bibitem{Eliashberg}  G.M. Eliashberg, \textit{Interactions between electrons and lattice vibrations in a superconductor}, Sov. Phys. JETP {\bf  11} (1960) 696
\bibitem{Bessorensen} D.R. Bes and R.A. Sorensen, {\it The pairing-plus-quadrupole model}, Adv. Nucl. Phys.  2 (1969) 129  
\bibitem {Idini_tesi}  A. Idini, Ph.D. thesis, University of Milano (2012), unpublished. \url{http://dx.doi.org/10.13130 \%2Fidini-andrea_phd2013-02-05}
\bibitem {Fetter} A. L. Fetter and J.D. Walecka,  {\it  Quantum theory of many-particle systems}, Mc Graw-Hill, New York (1971)
\bibitem{Terasakietal2002} J. Terasaki, F. Barranco, R.A. Broglia, E. Vigezzi and  P.F. Bortignon,
{\it Solution of the Dyson equation for nucleons in the superfluid phase}, Nucl. Phys. A {\bf 697} (2002) 127
\bibitem{Idinietal2012} A. Idini, F. Barranco and E. Vigezzi, {\it Quasiparticle renormalization and pairing correlations in spherical superfluid nuclei }, 
Phys. Rev. C  {\bf 85}, 014331 (2012)
\bibitem{Schuck} F. Barranco, P.F. Bortignon, R.A. Broglia, G. Col\`o, P. Schuck, E. Vigezzi and X. Vi\~nas, {\it Pairing matrix 
elements and pairing gaps with bare, effective and induced interactions}, Phys. Rev.  C {\bf 72}, 054314 (2005)
\bibitem{Schrieffer1964}  J.R.  Schrieffer, {\it Superconductivity} (Benjamin, New York, 1964)
\bibitem{BCS50} L.N. Cooper, {\it Remembrance of superconductivity past,}  in  {\it BCS: 50years}, Ed. by L.N. Cooper and  D. Feldman, World Scientific (2010) p. 3.
\bibitem{Cooper1956} L.N. Cooper, {\it Bound electron pairs in a degenerate Fermi gas}, Phys. Rev. {\bf 109}, 1189 (1956)
\bibitem{Yang1962} C.N. Yang {\it Concept of off-diagonal long-range order and the quantum phase of liquid He and
of superconductors}, Rev. Mod. Pays. {\bf  34}, 694 (1962)  
\bibitem{Ange} V. Ambegaokar, {\it The Green's function  method},  in {\it Superconductivity }, Ed. R.D. Parks,  Vol.  I,. 
Marcel Dekker Inc., New York, 259  (1969) 
\bibitem{Pais} A. Pais, {\it Inward bound}, Oxford  Univ. Press, Oxford (1986)
\bibitem{Lamb} W. Lamb and R. Retherford,  Phys. Rev. {\bf 72}, 241 (1947) 
\bibitem{Kroll} N.M. Kroll and W. Lamb, Phys. Rev. {\bf 75}, 388 (1946)
\bibitem{Bethe} R.P. Feynman, {\it Space-time approach to quantum electrodynamics}, Phys. Rev. 
{\bf 76}, 769 (1949)
\bibitem {Physica_scripta}  R. A. Broglia,  P. F. Bortignon,  F. Barranco,  E. Vigezzi,  A. Idini, and G. Potel,
{\it Unified description of structure and reactions: implementing the Nuclear Field Theory program}, Phys. Scr. \textbf{91} 063012 (2016)
\bibitem {11Be} F. Barranco, G. Potel, R.A. Broglia and E. Vigezzi,   {\it Structure and reactions of $^{11}$Be:
many-body basis for single-neutron halo},  to be published.  	arXiv:1702.01207 [nucl-th].
\bibitem{Mattuck} R.D. Mattuck, {\it A guide to Feynman diagrams in the many-body problem},
Dover, New York (1976)
\bibitem{Bes1977}D.R. Bes, G.G. Dussel, R.P.J. Perazzo and H.M. Sofia,  {\it The  renormalization of 
single-particle states in  nuclear field theory}, Nucl. Phys. A {\bf 293} (1977) 350
\bibitem{Bernard} V. Bernard and N.Van Giai,
 {\it Single-particle  and collective nuclear states 
 and the Green's function method}, Proc. Int. School of Physics E. Fermi,
 Course LVII, Eds. R.A. Broglia, R.A. Ricci and C.H. Dasso, 
 North Holland, Amsterdam (1981), p.437
 \bibitem{Guazzoni} P.  Guazzoni et al, {\it
$^{118}$Sn  levels studied by the  $^{120}$Sn (p, t) reaction: high-resolution measurements, shell model, and 
distorted-wave Born approximation calculations}, Phys. Rev. C  {\bf 78} 064608 (2013) 
\bibitem{An}  H. An and C. Cai, \textit{Global deuteron optical model potential for the energy range up to 183 MeV} Phys. Rev. \textbf{C 73}, 054605 (2006)
 \bibitem{Penrose} O,. Penrose and L. Onsager, {\it Bose-Einstein condensation and liquid helium}, Phys. Rev. {\bf 104}, 576 (1956)
 \bibitem{Pastore:08} A. Pastore, F. Barranco, R. A. Broglia and E. Vigezzi, \textit{Microscopic calculation and local approximation of the spatial dependence of the pairing field with bare and induced interactions}, Phys. Rev. \textbf{C 78}, 024315 (2008)
  
\bibitem{Lopez:13} N. Lopez Vaquero, J. L. Egido and T. R. Rodriguez, \textit{Large--amplitude pairing fluctuations in atomic nuclei},
  Phys. Rev. C \textbf{88}, 064311 (2013)
  
\bibitem{Bayman:72} B. F. Bayman and C. F. Clement,
\textit{Sum rules for two--nucleon transfer reactions},
    Phys. Rev. Lett. \textbf{29}, 1020 (1972)
    
\bibitem{Broglia:72} R. A. Broglia, C. Riedel, and T. Udagawa,
  \textit{Sum rule and two--particle units in the analysis of two--nucleon transfer reactions},
  Nucl. Phys. A \textbf{184}, 23 (1972)
  \bibitem{Lanford:77}W. A. Lanford, 
    \textit{Systematics of two--neutron transfer cross sections near closed shells: A sum--rule analysis of $(p,t)$ strength on the lead isotopes.},
    Phys. Rev. C \textbf{16}, 988 (1977)

\end{thebibliography}
\end{document}